# The European Union general data protection regulation: what it is and what it means[*]

Chris Jay Hoofnagle[a], Bart van der Sloot[b] and Frederik Zuiderveen Borgesius[c,d]

[a]Schools of Information and of Law, University of California, Berkeley, CA, USA; [b]Tilburg Institute for Law, Technology, and Society (TILT), Tilburg Law School (NL), Tilburg, Netherlands; [c]Institute for Computing and Information Sciences (iCIS), Radboud University (NL), Nijmegen, Netherlands; [d]Institute for Information Law (IViR), University of Amsterdam, Amsterdam, Netherlands

**ABSTRACT**

This paper introduces the strategic approach to regulating personal data and the normative foundations of the European Union's General Data Protection Regulation ('GDPR'). We explain the genesis of the GDPR, which is best understood as an extension and refinement of existing requirements imposed by the 1995 Data Protection Directive; describe the GDPR's approach and provisions; and make predictions about the GDPR's implications. We also highlight where the GDPR takes a different approach than U.S. privacy law. The GDPR is the most consequential regulatory development in information policy in a generation. The GDPR brings personal data into a detailed regulatory regime, that will influence personal data usage worldwide. Understood properly, the GDPR encourages firms to develop information governance frameworks, to in-house data use, and to keep humans in the loop in decision making. Companies with direct relationships with consumers have strategic advantages under the GDPR, compared to third party advertising firms on the internet. To reach these objectives, the GDPR uses big sticks, structural elements that make proving violations easier, but only a few carrots. The GDPR will complicate and restrain some information-intensive business models. But the GDPR will also enable approaches previously impossible under less-protective approaches.

## 1. Introduction

'Personal data is the new oil of the internet and the new currency of the digital world.'[1]

Suppose one bought into the metaphor of data as the new oil. One would want this new oil handled carefully. From extraction to disposal, all of its treatments would be planned carefully and executed by trained experts. One would want its extraction governed by a permit process, its uses managed to ensure it was not wasted, its storage secure, its disposal environmentally sound. One would want its externalities internalized and stakeholder interests considered.

The European Union's General Data Protection Regulation ('GDPR')[2] faithfully executes the implications of the oil metaphor, despite the metaphor's poor fit. The GDPR presumes that personal data are important, so much so that every aspect of interacting with data requires careful planning.

In this paper, we explain the GDPR approach to lawyers and academics, whether they are privacy and EU law specialists or not. We explain the GDPR's normative roots in multiple constitutional documents, detail its most important provisions, and tie these provisions to the short and medium-term strategic goals of the GDPR. We also highlight differences and similarities when comparing the GDPR to U.S. privacy law.

The GDPR has been law since 2016, but did not enter most lawyers' attention until 2018, when its provisions became enforceable.[3] In fact, much of the GDPR's requirements were reflected in an earlier law – the Data Protection Directive – which had poor enforcement and compliance. The GDPR awakened lawyers and the business community because it calls for minimum 8-figure fines and creates both internal and external mechanisms to bolster enforcement efforts.

As a result, the GDPR is the most consequential regulatory development in information policy in a generation. The GDPR brings personal data into a complex and protective regulatory regime. That said, the ideas contained within the GDPR are not entirely European, nor new. The GDPR's protections can be found – albeit in weaker, less prescriptive forms – in U.S. privacy laws and in Federal Trade Commission settlements with companies.[4]

To get to the GDPR, some level-setting is in order. First, one should not underestimate the commitment to data protection in Europe. The GDPR implements constitutional commitments, ones that are deep and occupy a central place in the self-conception of a new, information age political body. As one of the drafters of the Charter of Fundamental Rights of the European Union, Stefano Rodotà, explained,

> The fundamental right to personal data protection should be considered a promise just like the one made by the king to his knights in 1215, in the Magna Charta, that they would not be imprisoned or tortured illegally – 'nor will go upon him nor send upon him.' This promise, the habeas corpus, should be renewed and shifted from the physical body to the electronic body. The inviolability of the person must be reconfirmed and reinforced in the electronic dimension, according to the new attention paid to the respect for the human body (…).[5]

These commitments germinated long before the rise of contemporary Silicon Valley data companies but have only intensified as such companies have gained dominance.

---

[2]Regulation (EU) 2016/679 of the European Parliament and of the Council of 27 April 2016 on the protection of natural persons with regard to the processing of personal data and on the free movement of such data, and repealing Directive 95/46/EC (General Data Protection Regulation), OJ 2016 L 119/1 <http://ec.europa.eu/justice/data-protection/reform/files/regulation_oj_en.pdf> (hereafter, 'GDPR').

[3]GDPR art 99(2): 'It shall apply from 25 May 2018.'

[4]U.S. credit reporting laws have use limitations; communications laws regulate collection, use and sale of user data; the videotape privacy protection act establishes deletion requirements; credit reporting and cable and satellite providers must provide data subject access; and so on.

[5]S Rodotà, 'Data Protection as Fundamental Human Right,' in S Gutwirth, Y Poullet, P De Hert, C de Terwangne, and S Nouwt (eds), *Reinventing Data Protection?* (Springer, 2009).

To make the electronic body inviolable, the GDPR covers an immense landscape of potential informational problems. The GDPR attempts to answer information questions ex ante. Even remote, edge-case hypotheticals about data can be answered in the GDPR framework, with varying degrees of satisfaction.

Second, laws such as the EU's GDPR differ in construction from most U.S. regulatory text. The GDPR's text is vague in some places and speaks at the level of aspirational principle. Parts of the GDPR could be characterized as 'principles-based regulation'.[6] The GDPR's provisions are supplemented with even more indeterminate 'recitals.'[7] Such text flummoxes U.S. lawyers because of its lack of specificity.

Third, the difference in construction leads to a practical consequence: whereas in the U.S., interactions with regulators typically mean that enforcement is afoot, in the E.U. context, colloquy with regulators is a routine rite in the compliance process. U.S. lawyers have fretted about perfect compliance, but in reality, European regulators rarely expect such compliance, nor will they impose 8-figure liability for small imperfections. As we explain below, massive liability will also be keyed to serious wrongdoing rather than accident or simple noncompliance.

This paper does not aim to give detailed analyses of each GDPR provision. Rather, we focus on big themes, and often provide rough summaries of provisions, leaving out details that could be important in legal practice. Lawyers who apply the GDPR must, of course, consult the GDPR itself, and related guidance documents and case law.[8]

## 1.1. The GDPR's strategic implications

Throughout these sections, we discuss the strategic implications of the GDPR. We introduce eight key implications briefly here. First, the GDPR can be seen as a data governance framework. The GDPR encourages companies to think carefully about data and have a plan for the collection, use, and destruction of the data. The GDPR compliance process may cause some businesses to increase the use of data in their activities, especially if the companies are not data-intensive, but the GDPR causes them to realize the utility of data. Other businesses will use GDPR as an opportunity to more accurately evaluate the value of their data, converting the data to a strategic asset, on the same level as companies view their patent portfolio or copyrights.

Second, the GDPR attempts to put privacy on par with the laws that companies take seriously – antitrust and foreign corrupt practices law. Prior to the GDPR, large data companies faced low fines, often less than these companies pay a single entry-level engineer in a year.[9] The GDPR's enforcement mechanisms, penalty determination, expanded security incident notification, and procedural requirements that have the effect of documenting deviations from the law, will deter executives from seeing privacy violations as parking tickets. In addition, the GDPR's expanded view of what constitutes a 'breach' means that companies will file many more notices of security incidents.[10] Since the adoption of the GDPR, privacy and personal data are being discussed at the highest levels in companies. Many companies have revised their data practices, and take, for the first time, a professional approach to handling personal data.

---

[6]See R Baldwin, M Cave, and M Lodge, *Understanding Regulation: Theory, Strategy, and Practice* (2nd edn, Oxford University Press, Cambridge, 2011) 303; FJ Zuiderveen Borgesius, *Improving privacy Protection in the Area of Behavioural Targeting* (Kluwer Law International, 2015) 259–63.

[7]The Court of Justice of the European Union sometimes refer to recitals in data protection cases. See, e.g. Case C-131/12 *Google Spain SL, Google Inc. v Agencia Española de Protección de Datos (AEPD), Mario Costeja González* [2014] ECLI:EU: C:2014:317.]; See generally on the role of recitals T Klimas and J Vaičiukaitè, 'The Law of Recitals in European Community Legislation' (2008) 15 ILSA Journal of International & Comparative Law 3.

[8]A few commentaries on the GDPR have been published in English, such as D Rücker and T Kugler, *New European General Data Protection Regulation* (C.H. Beck Hart Nomos, 2018); P Voigt and A Von dem Bussche, *The EU General Data Protection Regulation (GDPR)* (Springer, 2017); European Agency for Fundamental Rights, 'Handbook on European Data Protection Law' (2018 edition) (Publications Office of the European Union, 2018). Several other teams are busy on article-by-article commentaries, including Christopher Kuner, Lee A. Bygrave, and Christopher Docksey (Oxford University Press, 2019) and Franziska Boehme and Mark Cole (2019).

Third, the GDPR requires protections to follow data. Companies that use personal data must vet service providers and impose contractual limits on data use. As a result, the GDPR has diffused far beyond first parties, to the many third parties involved in some kind of data services. Any large multinational requires GDPR compliance, thus, a fantastic number of service providers to such companies must make contractual commitments on data use, security, breach notification, and data retention. These service providers see requirements to comply with the GDPR as a form of economic coercion coming from their business associates rather than government coercion.

Fourth, the GDPR structurally elevates privacy officials within companies, giving them a tenure-like right but also responsibilities that transcend loyalty to the firm. The role of the Data Protection Officer, described in more detail below, will grease the wheels of enforcement, because these employees will be tasked with documenting what actually happens with data. Data depredations will be more difficult to hide. In enforcement proceedings, the accounting that companies have to undertake will provide the proof of non-compliance.

Fifth, the GDPR is constitutionally skeptical of U.S. lawyers' favorite tool: consent, particularly of the low-quality or 'take it or leave it' variety.[11] The GDPR's architects realized that if low-voluntariness consent could justify data activities, the GDPR would just become another exercise in clicking 'I agree' to unread, unnegotiable terms. The GDPR requires high-quality consent, on par with important life decisions, such as consent to medical treatment. In many contexts, the burdens the GDPR places on consent make consent impossible as mechanism to make data uses legal. Moreover, many rules in the GDPR are not waivable, and continue to apply after somebody has consented to data use.[12]

Sixth, as governments and technology companies dream of removing humans from decision-making processes, the GDPR makes a clear policy preference for human-in-the-loop systems. This happens as big data advocates have waived off concerns about erroneous data, arguing that large enough datasets make analysis possible with low-accuracy data. The GDPR tacks the other direction. GDPR's individual correction and remediation rights double down on the idea that personal data are important and should be accurate.

Seventh, the GDPR favors first-party relationships between data subjects and companies. Third party advertising and content networks have been the big winners online, but the GDPR recognizes the pathologies and misalignment flowing from third party data relationships. For instance, third party marketing companies exercise a 'tech tax' on publishers, leaving them with just $0.40 on each dollar spent on online advertising.[13] While the GDPR burdens third party data uses, it sees many controller uses of data as 'legitimate.'[14] First-party-favorable incentive structures could reshape internet commerce. For instance, data 'infomediaries,' companies that broker personal information on behalf of a data subject,[15] could become a practicable business model under the GDPR. Under less restrictive regimes, companies find it cheaper to buy personal data from third parties rather than the data subject.

---

[9]For instance, the CNIL levied its maximum fine against Facebook in 2017 for tracking users for advertising purposes. The fine was 150,000. See section 7 on enforcement.
[10]Reports of data breaches to the UK's Information Commissioner's Office have quadrupled in the months that the GDPR became enforceable. M Schwartz, 'Under GDPR, Data Breach Reports in UK Have Quadrupled,' *Bank Info Security* (25 July 2018) <https://www.bankinfosecurity.com/under-gdpr-data-breach-reports-in-uk-have-quadrupled-a-11249?>.
[11]C Hoofnagle, 'Designing for Consent' (2018) 7(2) Journal of European Consumer and Market Law 162. See in more detail on consent: Section 4.2 of this paper.
[12]For instance, the fair information principles from article 5 continue to apply after consent. See Section 4.1.

Eighth, there will be a period of experimentation, opportunism and mistake. When the GDPR became enforceable in May 2018, services barraged users with inane, widely-criticized privacy policies. Practitioners know that the real activity was 'under the hood' and substantial. The GDPR required extensive auditing of data collection and use; in the process, companies had to have a hard look at existing practices and the reliability of vendors. Many consent requests by companies were probably not in compliance with the GDPR. Companies 'in the sights' of the GDPR are likely to anchor with superficially acceptable practices, in the knowledge that enforcers will target them whether or not they fully embrace the law. We foresee an extended tussle between authorities and large companies such as Google and Facebook that involves positioning, anchoring, and other gamesmanship intended to blunt the GDPR's effects.

### 1.2. Roadmap

The GDPR's normative roots are deep. Section 2 below provides the historical background and political context of European data protection law and the GDPR. We recount the high-level commitments to privacy in its various forms, and how the EU's 1995 Data Protection Directive ('Directive') set the stage for the GDPR. In fact, companies in compliance with the Directive will find an evolution in privacy protections rather than a revolution. Yet few companies complied because of the relatively paltry fines in the Directive – fines that the GDPR greatly strengthens.

Sections 3–7 discuss the GDPR's contours, highlighting the issues of most concern to businesses. These include the GDPR's fines and enforcement mechanism, its presumption that data activities are illegal unless they have some sound basis, the broad definitions, its extraterritoriality, and the rights of data subjects. The GDPR covers a huge landscape of data activities; no U.S. information law is as broad and ambitious.

## 2. Background to the GDPR

Europe has long recognized privacy explicitly as a human right.[16] Europeans' commitments go beyond the home, the focus of so much U.S. law, to include protections for family life, communications, reputation,[17] and with the rise of the information age, for privacy in the context of data processing.

---

[13]R Benes, 'Why Tech Firms Obtain Most of the Money in Programmatic Ad Buys' *eMarketerPro* (16 April 2018) (reporting on J Mcdonald, 'Global Ad Trends, March 2018: Threats to digital advertising,' *Warc*, (March 2018)).
[14]See Section 4.2, and GDPR art 6(1)(f).
[15]Bethany L. Leickly, 'Intermediaries in Information Economies' (PhD dissertation, Georgetown University, 2004).
[16]In this paper, we use the phrases 'human right' and 'fundamental right' interchangeably. See on the difference: G Gonzalez Fuster, *The Emergence of Personal Data Protection as a Fundamental Right* (Springer, 2014) 164–6.

While U.S. lawyers may refer broadly to 'privacy' or to 'information privacy', European law discusses information privacy as 'data protection.'[18] In Europe, data protection is increasingly seen as separate from the right to privacy. Data protection focuses on whether data is used fairly and with due process[19] while privacy preserves the Athenian ideal of private life.[20]

The 1970s was a key point of privacy divergence between the U.S. and Europe. The U.S. articulated *Fair Information Practices* ('FIPs'),[21] the building blocks of all information privacy laws, but applied them in a serious sense only to the government in the form of the Privacy Act of 1974,[22] and to the private sector only in the credit reporting sector.[23] Europe embraced the FIPs, deepened and expanded them, and applied them to all information processing – both 'vertically,' that is government to citizen, and 'horizontally,' that is, business to citizen. The U.S. followed a sectoral model, leaving many forms of information practices regulated only by the Federal Trade Commission's general consumer protection authority.

By 1990, the European Commission feared that diverging national data protection laws would hinder the internal market in the EU.[24] That year, it published a proposal for a Data Protection Directive. After five years of negotiations, the final Data Protection Directive was adopted in 1995.[25] The Directive laid down an omnibus regime based on the FIPs, which applied to most of the private and public sector (with

---

[17] Article 8 of the 1950 European Convention on Human Rights provides protection to private and family life, home, and communication. Convention for the Protection of Human Rights and Fundamental Freedoms, Article 8, 4 November 1950, 213 U.N.T.S. 222. Most national constitutions in Europe also protect privacy and related rights. See on constitutional protection of privacy in European countries: BJ Koops, B Newell, T Timan, I Škorvánek, T Chokrevski, M Galič, 'A Typology of Privacy,' (2017) 38(2) University of Pennsylvania Journal of International Law 483.

[18] Information privacy 'concerns the collection, use and disclosure of personal information', PM Schwartz and DJ Solove, *Information Privacy* (Aspen, 2009) 1. Data privacy and information privacy refer to roughly the same concept.

[19] G Gonzalez Fuster, *The Emergence of Personal Data Protection as a Fundamental Right* (Springer, 2014).

[20] Consider the distinction made in the Charter of Fundamental Rights of the European Union: 'Article 7 – Respect for private and family life: Everyone has the right to respect for his or her private and family life, home and communications.'

> Article 8 – Protection of personal data: Everyone has the right to the protection of personal data concerning him or her. 2. Such data must be processed fairly for specified purposes and on the basis of the consent of the person concerned or some other legitimate basis laid down by law. Everyone has the right of access to data which has been collected concerning him or her, and the right to have it rectified. 3. Compliance with these rules shall be subject to control by an independent authority.

Charter of Fundamental Rights of the European Union (OJ C 364 of 18 December 2000) (hereafter, the 'Charter') <http://www.europarl.europa.eu/charter/pdf/text_en.pdf>. The Charter was adopted in 2000, and was made a legally binding instrument by the Treaty of Lisbon in 2009. Treaty of Lisbon amending the Treaty on European Union and the Treaty establishing the European Community [2007] OJ C306/01.

[21] See Robert Gellman, 'Fair Information Practices: A Basic History' (April 2017) <https://papers.ssrn.com/sol3/papers.cfm?abstract_id=2415020>; CJ Hoofnagle, 'The Origin of Fair Information Practices: Archive of the Meetings of the Secretary's Advisory Committee on Automated Personal Data Systems (SACAPDS)' (July 2014) <https://papers.ssrn.com/sol3/papers.cfm?abstract_id=2466418>; CJ Bennett, *Regulating Privacy: Data Protection and Public Policy in Europe and the United States* (Cornell University Press, 1992).

[22] Privacy Act of 1974 5 U.S.C. 552a (2017).

[23] Fair Credit Reporting Act of 1970 15 USC 1681 et seq (2017).

[24] The European Commission proposes and enforces legislation and implements policies and the EU budget. It can be seen as the EU executive power. See European Commission, 'The European Commission's Priorities' <https://ec.europa.eu/commission/index_en>.

[25] On the history of the Directive see: A Newman, *Protectors of Privacy: Regulating Personal Data in the Global Economy* (Cornell University Press, 2008).

exceptions to the latter).[26] The Directive required member states to enact implementing legislation.[27]

Problems quickly emerged with the Directive. The Directive did not fully harmonize national privacy laws, and even within Europe, countries behaved opportunistically to court big tech with signals of weak enforcement and advantageous tax schemes. (Some speak of data protection law's 'loophole' in Ireland.[28]) Even among countries committed to privacy, enforcement was limp wristed, with the French fining Facebook a mere 150,000 Euros in 2017.[29] This enforcement gap left Europe with a reputation of a region with rules but no real policing while the U.S. was seen as not being rule bound yet it had the Federal Trade Commission on the enforcement beat.[30]

The GDPR is the EU's attempt to address these and other shortfalls. It did so in a process completely unlike U.S. legislative efforts. European policy makers started a process that involved a multitude of expert consultation and deep sophistication about how information practices can be manipulated to evade regulatory goals.

Consultation began in 2009[31] and the European Commission published a proposal text in 2012.[32] Two years later, the European Parliament adopted a compromise text, based on almost 4,000 proposed amendments.[33] The Council of the European Union published its proposal for the GDPR in 2015, to start negotiations with the European Parliament.[34] In December 2015, the Parliament and Council reached agreement on the text of the GDPR. The GDPR was officially adopted in May 2016, and applies since May 2018.[35]

Now that the GDPR is enforceable, its interpretation is entrusted to the courts, combined with persuasive, albeit non-binding, interpretation by the newly created European Data Protection Board. The Court of Justice of the European Union (CJEU) is the highest authority on the interpretation of EU law.[36] The CJEU has delivered a number of dramatic,

---

[26]Directive 95/46/EC of the European Parliament and of the Council of 24 October 1995 on the protection of individuals with regard to the processing of personal data and on the free movement of such data (hereafter the 'Directive'). Some parts of the public sector are outside the scope of the Directive (see Article 3(2) and Article 13). Some data processing practices in the private sector are partially exempted; processing for purely personal purposes (Article 3(2)), and for journalistic purposes (Article 9).

[27]Consolidated Version of the Treaty on the Functioning of the European Union [2012] OJ C326/47 (TFEU) art 288 <http://eur-lex.europa.eu/legal-content/EN/TXT/PDF/?uri=CELEX:12012E/TXT&from=EN>.

[28]J Albrecht, '#EUdataP State of the Union,' (speech at the Chaos Communication Congress, 2013) <http://www.janalbrecht.eu/fileadmin/material/Dokumente/30C3-JPA-EUdataP.pdf>.

[29]CNIL, 'Facebook Sanctioned for Several Breaches of the French Data Protection Act,' May 16, 2017 <https://www.cnil.fr/en/facebook-sanctioned-several-breaches-french-data-protection-act>. See also Section 7 on enforcement.

[30]Kenneth Bamberger and Deirdre K Mulligan, *Privacy on the Ground* (MIT Press, 2015).

[31]European Commission, 'Commission Staff Working Paper, Impact Assessment Accompanying the document Regulation of the European Parliament and of the Council on the protection of individuals with regard to the processing of personal data and on the free movement of such data (General Data Protection Regulation) and Directive of the European Parliament and of the Council on the protection of individuals with regard to the processing of personal data by competent authorities for the purposes of prevention, investigation, detection or prosecution of criminal offences or the execution of criminal penalties, and the free movement of such data' (25 January 2012) 8.

[32]The European Commission is the EU's executive arm. See European Commission, 'The European Commission's priorities' (website) <https://ec.europa.eu/commission/index_en>.

[33]LIBE Compromise, proposal for a Data Protection Regulation (this paper refers to the unofficial Consolidated Version after LIBE Committee Vote, provided by the Rapporteur, General Data Protection Regulation, 22 October 2013. The European Parliament is an EU body with legislative, supervisory, and budgetary responsibilities. It is directly elected. <http://www.europarl.europa.eu/portal/>

[34]The Council of the European Union consists of government ministers from each EU country, according to the policy area to be discussed. European Council, 'Council of the European Union' <http://www.consilium.europa.eu/en/home/>.

[35]GDPR art 99(2): 'It shall apply from 25 May 2018.'

[36]National judges in the EU can, and in some cases must, ask the Court of Justice of the European Union if they are unsure how to interpret EU rules. For instance, if a national judge does not know how to interpret a national provision that is

pro-privacy decisions, including striking down the data retention mandates,[37] striking down the US–EU Safe Harbor agreement,[38] and granting people, under certain conditions, a 'right to be forgotten.'[39] Thus information industries face a more hostile judicial environment than in the U.S.

Like the Directive from 1995, the GDPR's first article stresses that the GDPR has a dual goal of promoting the free flow of personal data within the EU (to help businesses), and protecting people and their personal data.[40] Yet, the GDPR emphasizes the latter goal. The GDPR sets normative preferences in tension with information-intensive industry practices, particularly those performed by third parties. The GDPR lies in great tension with big data and machine learning business models, at least in their current form.

## 3. When does the GDPR apply?

### 3.1. Personal data and processing

The GDPR's architects studied information industry practices and even academic literature on cutting-edge identification techniques. For this and other reasons, the GDPR has extraordinarily broad scope in every dimension. Two threshold definitions are 'personal data,' and the information activities considered 'processing.'

All privacy laws define covered data. The GDPR sets a low bar, defining 'personal data' as 'any information relating to an identified or *identifiable* natural person ("data subject"); an identifiable natural person is one who can be identified, directly or indirectly (…)'.[41] Thus, the GDPR's concept of personal data is much broader than personally identifiable information such as names or addresses. In brief, every datum that identifies a person or could identify a person in the future is personal data. Public and non-sensitive information can also fall within the scope of 'personal data,' as do pseudonymous identifiers,[42] IP addresses, tracking cookies, and similar data. The GDPR also added 'location data' and 'online identifiers' as examples of identifiers in the GDPR's personal data definition.[43]

The GDPR applies when 'personal data' are 'processed'. The GDPR defines 'processing' as 'any operation or set of operations which is performed on personal data or on sets of personal data, whether or not by automated means (…)'.[44] This includes activities such as collecting, storing, disclosing, and erasing data. Consequently, practically everything that can be done with personal data will be considered to be 'processing'.

Taken together, these key definitions mean that, in principle, an organization processes 'personal data' within the meaning of the GDPR, whenever that organization touches data

---

based on an EU directive, the national judge must ask advice from the Court of Justice of the European Union (Article 19 (3)(b) of the Treaty on European Union, consolidated version).
[37]Joined Cases C-293/12 and C-594/12 *Digital Rights Ireland Ltd v Minister for Communications, Marine and Natural Resources and Others and Kärntner Landesregierung and Others* [2014] ECLI:EU:C:2014:238.
[38]Case C-362/14 *Maximillian Schrems v Data Protection Comr.,* [2015] ECLI:EU:C:2015:650.
[39]Case C-131/12 Google Spain SL, Google Inc. v Agencia Española de Protección de Datos (AEPD), Mario Costeja González [2014] ECLI:EU:C:2014:317.
[40]GDPR art (1). The GDPR 'lays down rules relating to the protection of natural persons with regard to the processing of personal data and rules relating to the free movement of personal data.'
[41]GDPR art 4(1).
[42]GDPR rec 26.
[43]The EU also adopted rules to safeguard the free flow of non-personal data: Regulation (EU) 2018/1807 of the European Parliament and of the Council of 14 November 2018 on a framework for the free flow of non-personal data in the European Union.
[44]GDPR art 4(2). GDPR recommends pseudonymization as a security measure. GDPR art 4(5). See also rec 28 and 29.

that relate to an individual, whether the data are public or private, sensitive or non-sensitive, directly or indirectly identify a person, and whether identification is possible now or in the future.

### 3.2. Who is accountable for upholding the GDPR requirements?

The four most important actors in the GDPR are 'data subjects', 'controllers', 'processors', and 'Data Protection Authorities'.[45] 'Data subjects' are people – the natural persons whose personal data are processed.[46] 'Controllers' are those who determine the purposes and the means of processing of personal data – companies for example.[47] 'Processors' are entities that do something with personal data on behalf of controllers;[48] in such case, there is a clear hierarchy. For example, if company Y gathers and analyzes survey data on the customers of company X, as instructed by company X, company X is the controller and company Y the data processor. If two organizations work together in determining why and how personal data will be processed, they will be seen as joint controllers and will share the regulatory burden and liability for errors and mistakes.

In the U.S. system, much is left to data subjects. They are supposed to read and critically evaluate privacy notices, and make choices in the marketplace based on this careful study.[49] Europeans reject that approach and place the clear burden of responsibility on controllers. Data processors, such as a data center or cloud provider, have to comply with a considerable proportion of the GDPR. If data processors violate the GDPR, in principle, the data controller will be considered responsible and liable.

The GDPR attempts to head off predictable principal-agent problems with processors. It requires that controllers ensure that processors are competent and responsible. To establish a chain of accountability, processors cannot subcontract without consent of the controller. The GDPR also specifies that if the subcontracted processor fails to fulfill its data protection obligations, the initial processor shall remain fully liable to the controller for the performance of that other processor's obligations.[50]

### 3.3. The GDPR's extraterritoriality

Personal data present several regulatory puzzles. For example, if companies can simply move information out of Europe to 'data havens' where no rules apply, the European privacy enterprise will fail. Thus, like the Directive did,[51] the GDPR imposes controls on personal data outside the EU. Another problem is that a small company with little revenue and few employees can still possess sensitive data on almost everyone in the world. Thus, the GDPR eschews traditional employee and revenue size thresholds for limiting coverage of the law.

---

[45] Apart from data controllers and data processors, there are several other positions, such as the recipient (GDPR art 4(9)), the third party (GDPR art 4(10)) and the representative (GDPR art 4(17)), but these positions fall outside the scope of this paper.
[46] GDPR art 4(1).
[47] GDPR art 4(7).
[48] GDPR art 4(8).
[49] D Solove, 'Privacy Self-Management and the Consent Dilemma' (2013) 126 Harvard Law Review 1879.
[50] See also GDPR art 29.
[51] See for the situation under the Directive, among others: C-191/15 *Verein für Konsumenteninformation v Amazon EU Sàrl* [2016] ECLI:EU:C:2016:612.

Obviously, a company that has offices or personnel in Europe will be subject to the GDPR.[52] For instance, the CJEU held that Google Spain is an establishment of Google Inc., and that some of the advertising activities aimed at the Spanish public were 'carried out in the context of the activities' of the Spanish establishment, so Google Inc. had to comply with EU law.[53]

But even if a company has no physical presence in Europe, the GDPR can apply.[54] For instance, offering goods or services, even free ones,[55] to Europeans can trigger the GDPR if it is 'apparent that the controller or processor envisages offering services to data subjects in one or more Member States in the Union.'[56] Simply making a website available is not enough, but using local language or currency may make it apparent that one is offering services to Europeans.[57] The basic philosophy of the GDPR is that when non-EU based organizations consciously process personal data of people in the EU, the GDPR will apply. (The nationality of the data subject is not relevant for the scope of the GDPR; the relevant criterion is whether people are in the EU.)

The GDPR's imposition of rules on third party trackers is among the clearest articulations of the strategy to limit such relationships. Even companies without a European establishment are subject to the GDPR if they use personal data for 'monitoring' the behavior of people in the EU.[58] For instance, if a behavioral targeting company tracks the browsing behavior of somebody in the EU. Indeed, the GDPR's preamble suggests that the rule is written specifically for online tracking.[59]

Consequently, if an American company places tracking cookies on the computers of people in the EU, the GDPR will apply. As we will see, the GDPR also burdens such tracking by classifying it as 'high risk.' The cumulative burdens placed on third parties could be a blessing for first parties, particularly news organizations and publishers that have grown dependent on third party tracking. The GDPR treats first-party data uses more leniently, even recognizing many of such uses as 'legitimate,' whereas if these same functions are performed by third parties, many substantive and procedural requirements are imposed.[60]

To enhance regulatory oversight, the GDPR requires controllers to establish a 'representative' to serve as a point person in interactions with both data subjects and Data Protection Authorities (supervisory authorities, roughly comparable with the Federal Trade Commission in the U.S.).[61] Such representative is a natural or legal person that is based on EU territory.[62] The obligation to appoint a representative applies to governmental organizations and private organizations that either process personal data on a large scale or that process sensitive personal data, such as regarding medical conditions, political beliefs or sexual preferences.

---

[52]GDPR art 3(1).
[53]Case C-131/12 *Google Spain SL, Google Inc. v Agencia Española de Protección de Datos (AEPD), Mario Costeja González* [2014] ECLI:EU:C:2014:317. See also Article 29 Working Party, 'Update of Opinion 8/2010 on applicable law in light of the CJEU judgement in Google Spain' (WP179 update) 16 December 2015.
[54]GDPR rec 2.
[55]GDPR art 3(2)(a); See also GDPR recs 22–24.
[56]GDPR rec 23.
[57]ibid.
[58]GDPR art 3(2).
[59]GDPR rec 24.
[60]See GDPR art 6(1)(f) and Section 4.2 of this paper.
[61]GDPR art 4(17), 27(4).
[62]GDPR art 3(2); See also GDPR rec 80.

### 3.4. What are the exceptions from the application of the GDPR?

The unprecedented scope of the GDPR is limited in two respects. First, a plain read of the GDPR suggests that we are all violating the GDPR, all the time, in our personal lives when we 'process' the 'personal data' of our friends by emailing them or by 'tagging them' in a photo online. In principle, these are indeed covered activities.

To keep regulatory focus on companies and organizations instead of ordinary people, the GDPR exempts data activities for 'purely personal or household activity.'[63] The GDPR's preamble shows that 'correspondence and the holding of addresses, or social networking' is an example of a situation that falls under the household exception.[64] For example, if somebody sends an email to a friend in which he writes about his nephew, this will normally not fall under the GDPR. But the exemption is narrow. The Court of Justice of the European Union held that when the surroundings of a house are filmed by a video camera attached to the home, for the purpose of identifying burglars, this will not count as a 'purely personal or household activity'.[65]

Second, the GDPR does not regulate national security. National security[66] and the prevention and prosecution of criminal offences,[67] are largely outside the scope of EU competence.[68] National member states hesitate giving up such competences to the EU. The Police Directive, which entered into force at the same time as the GDPR, sets rules for data processing by law enforcement agencies, such as the police. The rules in this Directive allow for more limitations than the general framework provided by the GDPR.[69] The GDPR does apply to data processing by other governmental organizations, although it allows governments to adopt special regimes in their national laws.[70]

In addition, the GDPR recognizes that the fundamental right to data protection might come into conflict with fundamental rights and other public interests, such as governmental transparency, freedom of speech, and data processing for archiving purposes. In Europe, freedom of speech is a fundamental right *on par with* privacy,[71] whereas in the U.S., freedom of speech generally trumps privacy when the two values are in conflict. The GDPR's architects were concerned that U.S.-style freedom of speech arguments could eat the rules. For instance, the U.S. Supreme Court's decision in *IMS Health* suggested that data selling – of medical prescription records – in preparation of direct marketing was constitutionally protected.[72] Such a fulsome extension of constitutional rights to marketplace activities would obviate many protections.

---

[63]GDPR art 2(2)(c). There are also special rules for national identification numbers (art 87), the context of employment (art 88), obligations of professional secrecy (art 90), and religious associations (art 91).
[64]GDPR rec 18.
[65]C-212/13 *František Ryneš v. Úřad pro ochranu osobních údajů* [2014] ECLI:EU:C:2014:2428.
[66]See also: C-473/12, *Institut professionnel des agents immobiliers (IPI) v Geoffrey Englebert, Immo 9 SPRL, Grégory Francotte, intervening parties: Union professionnelle nationale des détectives privés de Belgique (UPNDP), Association professionnelle des inspecteurs et experts d'assurances ASBL (APIEA), Conseil des ministers* [2013] ECLI:EU:C:2013:715.
[67]See also Article 29 Working Party, 'Guidelines for Member States on the criteria to ensure compliance with data protection requirements in the context of the automatic exchange of personal data for tax purposes', 175/16/EN, WP 234, 16 December 2015.
[68]GDPR art 23; GDPR rec 73.
[69]Directive 2016/680/EU of the European Parliament and of the Council of 27 April 2016 on the protection of natural persons with regard to the processing of personal data by competent authorities for the purposes of the prevention, investigation, detection or prosecution of criminal offences or the execution of criminal penalties, and on the free movement of such data, and repealing Council Framework Decision 2008/977/JHA OJ 2016 L 119/89. See also P De Hert and V Papakonstantinou Vagelis, 'The New Police and Criminal Justice Data Protection Directive. A First Analysis' (2015) 7(1) New Journal of European Criminal Law 7.
[70]GDPR art 23; GDPR rec 73.
[71]See S Kulk and FJ Zuiderveen Borgesius, 'Privacy, Freedom of Expression, and the Right to Be Forgotten in Europe' in J Polonetsky, O Tene, and E Selinger (eds) *Cambridge Handbook of Consumer Privacy* (Cambridge University Press, 2018).

The GDPR requires member states to balance data protection and freedom of speech interests, and requires that some of the GDPR's rules do not apply when personal data are processed for academic artistic or literary expression.[73] The EU left it to the member states to strike the difficult balance between data protection and freedom of expression, because different member states have different traditions in this context. For example, the Netherlands tends to give more weight to freedom of expression than Spain. We can also expect more case law from the CJEU on the balance between freedom of speech and data protection.

The GDPR makes clear that the right to privacy and other fundamental rights, such as freedom of expression, generally have equal weight. But the GDPR does allow for limitations on a small number of its rules, especially the rights of individuals,[74] when national governments deem that limitation necessary in terms of safeguarding other fundamental rights or public interests.[75]

## 4. Lawful data activities under the GDPR

The U.S. has a sectoral privacy regulation system, leaving many data activities subject only to general consumer protection law. Furthermore, data collection and use are in general permissible unless a law declares them not to be. The European approach is opposite: virtually all data activities are regulated in some way such that collection and use is always restricted. On a high level, the GDPR sets forth four rules to make data processing lawful: controllers and processors must adhere to the FIPs; the controller must have a legitimate ground for processing personal data; the controller must have grounds for using sensitive data, which are broadly defined under the GDPR; and the controller must have a lawful mechanism for moving data outside the EU.

The rights and responsibilities imposed by lawful activities reveal many elements of GDPR strategy. Viewed negatively, the GDPR appears to be a big data business killer. Viewed positively, the GDPR imposes a data governance approach, one where companies thoughtfully design business strategy to use data responsively and parsimoniously.

### 4.1. Adherence to Europeanized FIPs

The Fair Information Practices (FIPs) – the centerpiece of the 1995 Directive – remain at the core of the GDPR.[76] This is the basis of our claim that those in compliance with the Directive are well positioned to meet the GDPR's enhancements. Taken together, the imposition of the FIPs serves several of the GDPR's strategic aims. The FIPs attempt to minimize data collection and use. In the abstract, the FIPs are an appealing set of substantive and procedural protections against the power of data intensive companies. But taken together, the FIPs, create barriers to big data driven business models.[77]

---

[72]*Sorrell v IMS Health Inc.* 564 US 552 (2011).
[73]GDPR art 85. See further Case C-73/07 *Tietosuojavaltuutettu v Satakunnan Markkinapörssi Oy, Satamedia Oy* [2008] ECLI: EU:C:2008:727. See also D Erdos, 'European Regulatory Interpretation of the Interface between Data Protection and Journalistic Freedom: An Incomplete and Imperfect Balancing Act?,' University of Cambridge Faculty of Law Research Paper No. 61/2015 <https://ssrn.com/abstract=2683471>
[74]See infra Section 6 on individual rights.
[75]GDPR arts 85–91.
[76]See GDPR art 5, which lists the 'principles relating to processing of personal data', giving the main data protection requirements in broad terms. GDPR art 5 is based on the Directive's art 6. See GDPR rec 9. See generally on the data protection principles: LA Bygrave, *Data Privacy Law: An International Perspective* (Oxford University Press, 2014).

The GDPR's data protection principles resemble those developed in the U.S.[78] The GDPR FIPs articulate six high-level principles that data controllers and processors must consider. The FIPs apply cumulatively – each must be fulfilled in order for the data processing to be legitimate.

First, the lawfulness, fairness, and transparency principle articulates data protection law's overarching norm: personal data must be 'processed lawfully, fairly and in a transparent manner in relation to the data subject.'[79] The lawfulness requirement is reasonably clear: personal data processing must be compliant with the GDPR and other laws. The fairness requirement[80] could be compared with the general good faith requirement in some legal systems.[81]

Second, the purpose limitation principle entails that personal data should only be collected for a purpose that is specified in advance, and that those data should not be used for incompatible purposes.[82] The purpose should be specific and concrete; vague and abstract purposes such as 'promoting consumer satisfaction', 'product development' or 'optimizing services' are prohibited. A specific purpose exists, for example, when a pizza delivery service asks for the consumer's address, to deliver the pizza.

In brief, to assess whether a new purpose is compatible with the original purpose, the controller should consider, for instance, the link between the original and new purposes, the context, the data subject's reasonable expectations, the data's nature and sensitivity, the consequences of the intended further processing for data subjects.[83] The GDPR does not consider further processing to be incompatible with the initial purposes, if the further processing happens for archiving purposes in the public interest, scientific or historical research purposes, or statistical purposes.[84] However, the GDPR requires extra safeguards for such further processing.[85]

Purpose limitation strikes at the heart of information-intensive industries, because companies so frequently find utility for data by using and repurposing the data in unforeseeable ways. Indeed, the very purpose of machine learning is to discover patterns not anticipated or even perceivable to people. While many big data enthusiasts have suggested that the purpose limitation principle is outdated and should be abolished, the GDPR retains the purpose limitation principle. Hence, on this point the EU made a political choice in favor of privacy and data protection.

---

[77]Some even argue that the GDPR represents a rejection of big data driven business models: Tal Zarsky, Incompatible: The GDPR in the Age of Big Data (2017) 47(2) Seton Hall Law Review 995.

[78]The FIPs can be recognized, for example, in U.S. Department of Health, Education & Warfare, Records and the Rights of Citizens (1973) <www.justice.gov/opcl/docs/rec-com-rights.pdf>, in the US Privacy Act of 1974, Pub. L. No. 93–579, 88 Stat. 1896 (codified at 5 U.S.C. § 552a (2012)). Similar principles are included, for instance, in OECD, 'Guidelines on the Protection of Privacy and Transborder Flows of Personal Data,' <http://www.oecd.org/sti/ieconomy/oecdguidelinesontheprotectionofprivacyandtransborderflowsofpersonaldata.htm>.

[79]GDPR art 5(1)(a).

[80]See also GDPR recs 39, 42, 57, 58, 60, 61, 62, 71. The EU Charter also says that personal data processing must happen 'fairly', Charter art 8(2).

[81]See the good faith principle in the Draft Common Frame of Reference: '[t]he expression "good faith and fair dealing" refers to a standard of conduct characterised by honesty, openness and consideration for the interests of the other party to the transaction or relationship in question.' Study Group on a European Civil Code, Principles, Definitions and Model Rules of European Private Law: Draft Common Frame of Reference, C Von Bar, E Clive and H Chulte-Nölke (eds) <https://www.law.kuleuven.be/personal/mstorme/DCFR.html>.

[82]The GDPR puts it as follows: personal data shall be 'collected for specified, explicit and legitimate purposes and not further processed in a manner that is incompatible with those purposes.' GDPR art 5(1)(b). The principle is also included in the Charter of Fundamental Rights of the European Union. Charter art 8(2). See also Article 29 Working Party, 'Opinion 03/2013 on purpose limitation', 00569/13/EN, WP 203, 2 April 2013.

[83]GDPR art 6(4); GDPR rec 50.

[84]GDPR art 5(1)(b).

[85]GDPR art 5(1)(b) and art 89.

Third, the data minimization principle holds that personal data should be 'adequate, relevant and limited to what is necessary in relation to the purposes for which they are processed.'[86] The preamble adds that '[p]ersonal data should be processed only if the purpose of the processing could not reasonably be fulfilled by other means.'[87] Only those data that are needed for the specific purpose may be obtained. Thus, the pizza delivery service should not collect data about people's religious or political views – after all, such data are not necessary for delivering the pizza. The data minimization principle thus prohibits collecting as much personal data as possible because the data could be useful in the future, in a way rejecting many big data business models.

Fourth, the accuracy principle requires that personal data are 'accurate and, where necessary, kept up to date.'[88] Data controllers must take 'every reasonable step (…) to ensure that personal data that are inaccurate, having regard to the purposes for which they are processed, are erased or rectified without delay.'[89] Thus, the accuracy principle does not always require full accuracy; it requires accuracy 'having regard to the purposes' for which personal data are processed.[90] Data controllers must proactively ensure appropriate accuracy, and must offer data subjects the possibility to correct data.[91]

Fifth, in addition to minimizing data, the GDPR tightly limits data storage.[92] The principle imposes a 'no longer than necessary' standard. The preamble adds that controllers should set, ex ante, time limits for planned erasure.[93] Thus, the pizza delivery service should not store customer addresses for unreasonably long periods. Deleting the address once the pizza has been delivered would be perfect. But the pizza place could also keep the address for a few months to save returning customers the time of dictating their address again.

Sixth, the integrity and confidentiality principle imposes data security responsibilities. Security must be 'appropriate' and protect against loss, destruction, damage and unlawful processing, thus *internal uses* of data can be security violations.[94] This will be explained in more detail in Section 5, regarding the responsibilities for data controllers. In sum, the FIPs form the core of the GDPR – the GDPR is FIPs on steroids.

---

[86]GDPR art 5(1)(c).
[87]GDPR rec 39.
[88]GDPR art 5(1)(d).
[89]GDPR art 5(1)(d). See also GDPR rec 71, concerning accuracy in the area of profiling.
[90]GDPR art 5(1)(d).
[91]GDPR art 16.
[92]'Personal data shall be … kept in a form which permits identification of data subjects for no longer than is necessary for the purposes for which the personal data are processed; personal data may be stored for longer periods insofar as the personal data will be processed solely for archiving purposes in the public interest, scientific or historical research purposes or statistical purposes in accordance with Article 89(1) subject to implementation of the appropriate technical and organisational measures required by this Regulation in order to safeguard the rights and freedoms of the data subject ("storage limitation")', GDPR art 5(1)(e).
[93]GDPR rec 39.
[94]GDPR art 5(1)(f).

## 4.2. The legal basis for processing personal data

Recall that Europe's privacy approach enshrines data protection as a fundamental right.[95] To implement this commitment, the GDPR specifies six legal justifications for data processing, which were copied almost verbatim from the Directive.[96] Data processing should, in addition to adhering to the FIPs, be based on one of six grounds. Roughly summarized, these grounds are: (1) the data subject has consented to the data processing, (2) the data processing is necessary for a contract with the data subject, (3) there is a law mandating the data processing (e.g. tax law requires companies to keep certain records), (4) data processing is necessary to protect the life of a data subject (e.g. the data subject is unconscious after a car accident, and the hospital needs to know from the data subject's family doctor whether the data subject uses certain medication) (5) data processing happens for a public task (e.g. the tax office gathers certain data, such as people's tax returns, to fulfill its tasks), and (6) when the interests of the data controller prevail over the interests of the data subject. We focus on the most important for businesses: (1) consent, (2) contractual necessity, and (6) legitimate interests.[97]

Notice and choice dominates the U.S. approach to privacy.[98] As companies develop new uses of data, lawyers change privacy policies and users 'consent' to them, on pain of being denied services.[99] The U.S. approach has been called a 'successful failure' that portrays the market as a privacy-protecting mechanism while blame resides on individuals for failing to achieve privacy through their 'choices.'[100]

In contrast, the GDPR imposes substantive and procedural requirements on consent. Under the GDPR, consent must be freely given, specific, informed, unambiguous,[101] and revocable.[102] Each of these requirements has bite. Valid consent requires an unambiguous indication of wishes. Hence, companies cannot obtain valid consent by using an opt-out system in which failure to object is supposed to signify consent. Valid consent requires 'a statement' or 'a clear affirmative action' from the data subject.[103]

The GDPR emphasizes that consent must be genuinely 'freely given' to be valid. The 'freely given' requirement prohibits, in many circumstances, take-it-or-leave-it conditions regarding privacy.[104] Take-it-or-leave-it choices are ubiquitous: many services only allow people to use them if they accept being tracked for targeted marketing. And many websites use 'tracking walls', barriers that people can only pass if they agree to third party tracking.[105] Under which circumstances take-it-or-leave-it choices are still acceptable has to become clear from enforcement. The day the GDPR became enforceable, Max Schrems complained to Data Protection Authorities about the take-it-or-leave-it practices of Google, Instagram, WhatsApp, and Facebook.[106]

---

[95]Charter art 8(2). See also GDPR rec 40. See also: Case C-131/12 *Google Spain SL, Google Inc. v Agencia Española de Protección de Datos (AEPD), Mario Costeja González* [2014] ECLI:EU:C:2014:317, par 71.
[96]See Directive art 7.
[97]See on the six legal bases: Article 29 Working Party, 'Opinion 06/2014 on the notion of legitimate interests of the data controller under article 7 of Directive 95/46/EC' (WP 217) 9 April 2014.
[98]D Solove, 'Privacy Self-Management and the Consent Dilemma' (2013) 126 Harvard Law Review 1880.
[99]PM Schwartz, 'Internet Privacy and the State' (2000) 32 Connecticut Law Review 815.
[100]Gordon Hull, 'Successful Failure: What Foucault Can Teach Us about Privacy Self-Management in a World of Facebook and Big Data' (2015) 17(2) Ethics and Information Technology 89.
[101]GDPR arts 4(11), 6(1); GDPR recs 32, 33, 42, and 43. See also Article 29 Working Party, 'Guidelines on consent under Regulation 2016/679' (WP259 rev.01), Brussels, 10 April 2018.
[102]GDPR art 7(3).
[103]GDPR art 4(11). See also GDPR art 7.
[104]GDPR art 7(4). See also Article 29 Working Party 2012, 'Opinion 04/2012 on Cookie Consent Exemption', WP 194, 7 June 2012; Article 29 Working Party, 'Guidelines on consent under Regulation 2016/679' (WP259 rev.01), Brussels, 10 April 2018.

The 'specific' and 'informed' requirements prohibit the use of vague consent requests. For example, asking consent 'to use your data for commercial purposes' cannot lead to valid consent. Moreover, the GDPR prohibits burying a consent request in the small print of an End User License Agreement or similar document.[107] Revocable means there needs to be symmetry in enrollment and cancelation – revocation must be as easy as it was to sign up.[108] In sum, because of the GDPR's strict requirements for valid consent, consent is, in many cases, simply not practicable as a lawful basis for processing personal data.

Even if a controller demonstrates these consent elements, the GDPR limits the scope of consent to approval of processing. That is, consent does not waive other GDPR limitations, such as minimization, accuracy, or deletion responsibilities.[109]

Consent is further restraint for children in the context of online services. Children cannot give consent until they are 16, a major imposition on social network services such as Facebook that worry about whether younger users will use other systems. Member States may set a lower minimum consent age, but not lower than 13 years.[110] Finally, the GDPR places the burden to demonstrate the lawfulness of consent on the controller.[111]

Contractual necessity is a second, obviously necessary, basis for lawful processing. The GDPR allows processing 'necessary for the performance of a contract to which the data subject is party or in order to take steps at the request of the data subject prior to entering into a contract.'[112] The processing must be genuinely necessary to perform the contract for this basis to apply.[113] To illustrate: if somebody orders a pizza, the pizzeria can give the customer's address (a piece of personal data) to the delivery person, because the address is 'necessary' to deliver the pizza, and thus to perform the contract.

Taken together, the imposition of costs on gaining consent, and the ruination of its benefits caused many lawyers to tell clients not to use consent. Instead, lawyers' strategy focused on 'legitimate interests', which we turn to next.

---

[105]See FJ Zuiderveen Borgesius et al., 'Tracking Walls, Take-it-Or-Leave-it Choices, the GDPR, and the ePrivacy Regulation' (2017) 3 European Data Protection Law Review 353.
[106]NOYB, 'GDPR: noyb.eu filed four complaints over 'forced consent' against Google, Instagram, WhatsApp and Facebook' 25 May 2018, <https://noyb.eu/wp-content/uploads/2018/05/pa_forcedconsent_en.pdf>.
[107]The consent must be 'clearly distinguishable from the other matters, in an intelligible and easily accessible form, using clear and plain language.' GDPR art 7(2).
[108]GDPR art 7(3).
[109]Article 29 Working Party, 'Guidelines on consent under Regulation 2016/679' (WP259 rev.01), Brussels, 10 April 2018.
[110]GDPR art 8(1). For children below 16, the parent (or holder of parental responsibility) should authorize consent. The controller should make 'reasonable efforts' to verify that consent is given or authorized by the parent, taking into consideration available technology. GDPR art 8(2). See generally on children and consent to data processing M Macenaite and E Kosta, 'Consent for processing children's personal data in the EU: following in US footsteps?' (2017) 26(2) Information & Communications Technology Law 1. See for an overview of the age of consent in EU member states: I Milkaite and E Lievens, 'GDPR: updated state of play of the age of consent across the EU, June 2018', <https://www.betterinternetforkids.eu/web/portal/practice/awareness/detail?articleId=3017751>.
[111]GDPR art 7(1). See for an example of a system to register consent: <https://privacybydesign.foundation/irma-en/>.
[112]GDPR art 6(1)(b). See also GDPR recs 40 and 44.
[113]See Article 29 Working Party, 'Opinion 06/2014 on the notion of legitimate interests of the data controller under article 7 of Directive 95/46/EC', WP 217, 9 April 2014, 17.

The legitimate interests provision, also known as the balancing provision, is a catchall basis that attempts to recognize the multifarious reasons why personal data needs to be used.[114] In order to stop legitimate interests from becoming too broad of a basis, the GDPR uses a balancing test to compare the relative interests of the controller and data subject. In the words of the GDPR: processing is lawful if it is

> necessary for the purposes of the legitimate interests pursued by the controller or by a third party, except where such interests are overridden by the interests or fundamental rights and freedoms of the data subject which require protection of personal data … [115]

Many standard business functions fit under legitimate interests. For instance, website publishers can store IP addresses of website visitors for a brief period, if that is necessary for security or for fraud prevention.[116] And a pizza delivery service may store the customer's address, to send the customer a letter if the pizzeria introduces a new menu, or has a special offer. Storing a customer's name and address does not, in general, heavily infringe privacy, and the pizzeria has a legitimate interest in promoting its own business. The pizzeria would probably pass the balancing test prescribed by the legitimate interests provision.

Compared to the Directive, the GDPR sets forth more guidance than the Directive on the legitimate interests provision,[117] and imposes new requirements. For instance, legitimate interests must be explicitly disclosed,[118] and there is more emphasis placed on the rights of children.[119]

Some lawyers hope that many marketing practices, including online behavioral advertising, can be based on the legitimate interests provision.[120] Can you use the legitimate interests provision that broadly? Well, it depends.[121] Certain types of innocuous targeted marketing, for instance paper direct marketing for the controller's own products to existing customers, can be based on the legitimate interests provision. More information-intensive marketing, such as behavioral advertising, generally requires the data subject's prior consent.[122]

Moreover, there are specific rules (outside of the GDPR) for certain types of direct marketing. For instance, robocalls and email marketing are only allowed after the recipient's prior consent.[123] Prior consent of the recipient is also required for the use of tracking cookies and similar tracking technologies for behavioral advertising.[124]

---

[114]GDPR art 6(1)(f). See also GDPR recs 47–50.
[115]GDPR art 6(1)(f).
[116]See: CJEU, C-582/14 *Patrick Breyer v. Bundesrepublik Deutschland* [2016], ECLI:EU:C:2016:779. See on the legitimate interests provision also: CJEU, Valsts policijas Rīgas reģiona pārvaldes Kārtības policijas pārvalde v Rīgas pašvaldības SIA 'Rīgas satiksme', C-13/16, 4 May 2017.
[117]GDPR recs 37, 47 and 49.
[118]GDPR art 13(1)(d).
[119]GDPR art 6(1)(f).
[120]See, e.g. Hunton & Williams LLP, Centre for Information Policy Leadership GDPR Implementation Project, 'Recommendations for Implementing Transparency, Consent and Legitimate Interest under the GDPR,' 19 May 2017.
[121]See GDPR recs 10, 47, 48, and 49.
[122]Article 29 Working Party, 'Opinion 03/2013 on purpose limitation', WP 203, 2 April 2013, 46. See also FJ Zuiderveen Borgesius, 'Personal Data Processing for Behavioural Targeting: Which Legal Basis?', International Data Privacy Law 2015-5-3, p. 163–176.
[123]Directive 2002/58/EC of the European Parliament and of the Council of 12 July 2002 concerning the processing of personal data and the protection of privacy in the electronic communications sector, as amended by Directive 2009/136/EC of the European Parliament and of the Council of 25 November 2009 <https://eur-lex.europa.eu/eli/dir/2002/58/2009-12-19> (hereafter the 'ePrivacy Directive'), art 13.
[124]ePrivacy Directive art 5(3).

If a controller relies on the legitimate interests provision for targeted marketing, the GDPR grants people an absolute right to object to (opt out of) that marketing, even for innocuous direct marketing practices such as direct mail.[125] The GDPR has a separate provision that refers to Do Not Track-like systems: in the online context, 'the data subject may exercise his or her right to object by automated means using technical specifications.'[126] The Do Not Track standard should enable people to signal with their browser that they do not want to be tracked on the internet.[127]

### 4.3. Sensitive data

Sensitive data processing is another pain point for businesses. The GDPR both broadly defines sensitive data – termed 'special categories of personal data'[128]—and explicitly prohibits their use (subject to exceptions). Special categories of data include personal data revealing racial or ethnic origin,[129] political opinions,[130] religious or philosophical beliefs, or trade-union membership, data concerning health,[131] data concerning a natural person's sex life or sexual orientation, genetic data and the processing of biometric data for the purpose of uniquely identifying a natural person. The CJEU has adopted a wide interpretation of these classes. For example, when somebody mentioned on a website that a colleague injured her foot and worked part-time on medical grounds, the CJEU said that the this constituted disclosing personal data concerning health.[132] Americans might find this approach unreasonable, but on the Continent, the Nazi, Stasi, and Soviet uses of special category data are still in the minds of the living.

The problem for businesses is that sensitive data are valuable for ad targeting. Many marketers, especially in the U.S., frame their consumer segments in terms of racial or political identity. And the many exceptions to the ban on processing are impracticable or simply not suited for most businesses.

Although the GDPR prohibits the processing of sensitive data, there are several exceptions. The most felicitous is consent, but in addition to the GDPR's already high bar for consent, special categories data requires consent to be 'explicit.'[133] So strong was the GDPR's commitment to this rule that Member States may enhance it by eliminating the consent exception entirely.[134]

The GDPR allows processing of special categories of data where the 'data subject has manifestly made his or her personal data public'.[135] This exception applies, for instance, to data about politicians who publicly tell about their political opinions. The exception can also apply if somebody openly and explicitly stated that he or she is gay or has a certain medical condition, addressed at the public at large and intended to be accessible to everyone. Another exception concerns employment relationships; in that context businesses may use special category data in some circumstances.[136] Another exception covers processing necessary for preventive or occupational medicine.[137] The other exceptions are either irrelevant or poorly suited to business needs. The prohibition can be lifted, in certain circumstances, when the processing of sensitive data is necessary for:

- protecting the vital interests of the data subject or of another person when they are physically or legally incapable of giving consent.[138]
- the functioning of a church, trade union, political party, or similar non-profit organization.[139]
- for the establishment, exercise, or defense of legal claims, or whenever courts are acting in their judicial capacity.[140]
- for reasons of substantial public interest.[141]
- for a specific, substantial public interest, such as public health.[142]
- for archiving purposes in the public interest, scientific or historical research purposes or statistical purposes.[143]

---

[125]GDPR art 21; GDPR rec 47.
[126]GDPR art 21(5). See also GDPR rec 59.
[127]See generally on Do Not Track and European law: FJ Zuiderveen Borgesius, J Van Hoboken, K Irion, and M Rozendaal, 'An Assessment of the Commission's Proposal on Privacy and Electronic Communications,' Directorate-General for Internal Policies, Policy Department C: Citizen's Rights and Constitutional Affairs, June 2017 <https://ssrn.com/abstract=2982290>.
[128]GDPR art 9. The Data Protection Directive had a similar regime in art 8.
[129]Regarding data about racial origin, the preamble says that the EU does not accept theories that try to determine the existence of separate human races (GDPR rec 51).
[130]See on personal data revealing political opinions: GDPR rec 56.
[131]See on data relating to health: GDPR recs 35, 45, 52–54, 63, 65, 71, 75, 91. Health data can relate to both a person's physical and mental health, including the provision of health care services, which reveal information about his or her health status. GDPR art 4(15).
[132]C-101/01 *Bodil Lindqvist* [2003] ECLI:EU:C:2003:596.
[133]GDPR art 9(2)(a).
[134]ibid.
[135]GDPR art 9(2)(e).

Like the Directive, the GDPR contains a separate regime for personal data relating to criminal convictions and offences. In brief, only an official authority may process personal data relating to criminal convictions and offences or related security measures. A comprehensive register of criminal convictions may only be kept under the control of official authority.[144] This stands in stark difference from the U.S. landscape, where arrest records are open to public inspection and copying. Even where the suspect is never charged or not convicted, this information remains in the public record and is actively traded by companies.

### 4.4. Moving data outside Europe – cross border data transfers

The EU's strong rules for data could be made futile if information can simply be moved to a 'data haven' with fewer or no restrictions. Thus, the GDPR establishes a framework for international transfers of data.[145] The rules are detailed, but can be roughly summarized as follows. Transfer is allowed in two situations.[146]

---

[136]GDPR art 9(2)(b). GDPR rec 52.
[137]For example, when the processing of health data is necessary for the assessment of the working capacity of the employee, medical diagnosis, the provision of health or social care or treatment or the management of health or social care systems and services. GDPR art 9(2)(h).
[138]GDPR art 9(2)(c).
[139]GDPR art 9(2)(d).
[140]GDPR art 9(2)(f).
[141]The GDPR adds an additional requirement when it comes to legitimizing the processing of special categories of data (sensitive data). The processing should not only be in the public interest, but in the 'substantial' public interest. The GDPR does not specify what is to be regarded as substantial. GDPR art 9(2)(g).
[142]Examples include protecting against serious cross-border threats to health or ensuring high standards of quality and safety of health care and of medicinal products or medical devices. GDPR art 9(2)(i).
[143]GDPR art 9(2)(j). See also GDPR art 89.
[144]GDPR art 10.

First, the European Commission can evaluate other countries' legal regimes and find them adequate.[147] The Commission has approved only eleven jurisdictions: Andorra, Argentina, Canada, Faeroe Islands, Guernsey, Isle of Man, Israel, Jersey, New Zealand, Switzerland, and Uruguay.[148] Such approval is normally only given after long negotiations, in which the country in question updates its national data protection regime to a level that is nearly as protective as the EU-data protection rules.

Second, when the data are transferred to an organization that is based in a non-EU country for which there is no adequacy decision, this will only be deemed legitimate when the organization in question contractually guarantees that it will uphold, within its organization, a level of data protection that is similar to the GDPR, including all of the material and procedural safeguards. Consequently, the data transferred to an organization in a non-EU country will still be under a similar level of protection as when they would have stayed on EU-territory.[149]

For the U.S., which does not have the status of a country with 'adequate' protection, a special 'Safe Harbor' arrangement was in place. In short, companies from the U.S. from certain sectors were deemed to offer an adequate level of protection if they agreed to comply with the data protection principles.[150] But a single Austrian lawyer, Max Schrems, architected a legal challenge to the Safe Harbor that led to its invalidation. Indeed, at the time of writing, the most powerful companies in America are uncertain about how they will move personal data from Europe in the future.[151]

In invalidating the Safe Harbor, the CJEU, concerned about the Snowden disclosures,[152] made a broad ruling holding that widespread government access to communications data and the lack of effective remedies for such access violated the Charter of Fundamental Rights of the European Union.[153] Although ruling in the context of the Safe Harbor, the reasoning of the Court would suggest that the other mechanisms for transfer are vulnerable to challenge too.

Following the decision, the European Commission negotiated a new agreement with the U. S., the EU-U. S. Privacy Shield,[154] in which some of the concerns raised by the

---

[145] GDPR art 44. See also PM Schwartz and KN Peifer, 'Transatlantic Data Privacy,' (2017) 106 Georgetown Law Journal 115.
[146] For specific circumstances, an exception may apply; these exceptions are as strict and rigorous as those that apply to the processing of sensitive data. In addition, the exceptions may only be used for incidental, small scale data processing, such as when the data about one consumer is transferred in a singular instance. GDPR art 49.
[147] GDPR art 45.
[148] European Commission, 'Data Transfers Outside the EU' <https://ec.europa.eu/info/law/law-topic/data-protection/data-transfers-outside-eu_en>.
[149] GDPR arts 46–47.
[150] Commission Decision 2000/520/EC of 26 July 2000 pursuant to Directive 95/46 on the adequacy of the protection provided by the safe harbour privacy principles and related frequently asked questions issued by the US Department of Commerce. See also Article 29 Working Party, 'Opinion 01/2016 on the EU–U.S. Privacy Shield draft adequacy decision', 16/EN, WP 238, 13 April 2016. Article 29 Working Party, 'Statement of the Article 29 Working Party', 16 October 2015.
[151] The research for this paper was concluded on 16 January 2018.
[152] These adequacy decisions do not cover data exchanges in the law enforcement sector. For special arrangements concerning exchanges of data in this field: European Commission, 'Transfer of Air Passenger Name Record Data and Terrorist Finance Tracking Programme' <http://ec.europa.eu/justice/data-protection/international-transfers/pnr-tftp/pnr-and-tftp_en.htm>.
[153] Case C-362/14 *Maximillian Schrems v Data Protection Comr.*, [2015] ECLI:EU:C:2015:650.
[154] Commission Implementing Decision (EU) 2016/1250 of 12 July 2016 pursuant to Directive 95/46/EC of the European Parliament and of the Council on the adequacy of the protection provided by the EU-U.S. Privacy Shield (notified under document C(2016) 4176).

Court of Justice have been addressed. However, other points of concern remain. Still, for the foreseeable future, the Privacy Shield is the mechanism of choice for most U.S. businesses. At a high level, the Privacy Shield requires a company to commit to a GDPR-like level of protection, including all or its principles, prohibitions and material and procedural requirements. The transfer of personal data from Europe to the U.S. has been a problem since at least 20 years,[155] and will remain a problem, as it is not likely that the U.S. will adopt privacy rules that could lead to an adequacy status.

## 5. Responsibilities for data controllers and processors

The GDPR reaffirms the role of the data controller as the party responsible for data, and imposes stricter controls, duties, and even liability on processors. This came about because under the Directive, some companies attempted to avoid privacy responsibilities by declaring themselves to be 'processors,' while engaging in controller-like decision making. In addition, while the Directive granted a large role to the governmental supervisory organizations, Data Protection Authorities, for monitoring and enforcing the data protection rules, the GDPR places the primary burden for supervision and control on the data controllers themselves.

There are 9 obligations the data controller has to abide by, next to the principles discussed in Sections 3 and 4 of this paper, in order to be GDPR compliant. First, data controllers must keep detailed, accounting-like records of their processing activities.[156] Strategically, this obligation not only saddles controllers with ultimate responsibility for data processing;[157] it creates an enforcement roadmap for regulators. The onus lies principally on controllers to provide either a judge or a Data Protection Authority with detailed information to show that they have acted carefully and legitimately. The record must contain information such as which personal data is processed, why, whom has access to them, how long they are stored and which security measures have been adopted.

The record-keeping provisions depart from the strategy in the Directive. The Directive had a woefully unrealistic requirement that all controllers notify the Data Protection Authority when planning to engage in any wholly or partly automatic processing operation.[158] The Directive envisioned a permit-style approach enabling Data Protection Authorities to keep a register of data processing activities and to assess the proposed initiatives beforehand or intervene at an early stage.[159] Obviously this was unfeasible and the Directive allowed states to limit the notification requirement.[160] The GDPR eliminated this ex ante approach.

Second, the data controller must adopt a data protection policy. In it, the data controller specifies and gives reasons for the various choices made. Why has the controller gathered personal data; why so many data; for what purpose has it gathered those data; how will it ensure that the data are maintained correct and up to date; who has access to the data within the organization and why do these people need that access; etc.[161]

---

[155]See: D Heisenberg, Negotiating Privacy: The European Union, the United States, and Personal Data Protection (Lynne Rienner Publishers 2005).
[156]GDPR art 30.
[157]See also GDPR rec 82.
[158]Directive arts 18–19.
[159]Directive art 20.
[160]Directive art 19.

Third, the GDPR requires data controllers to be transparent, even when data subjects do not specifically request information.[162] The GDPR requires data controllers to provide information 'in a concise, transparent, intelligible and easily accessible form, using clear and plain language … '.[163] The GDPR specifies a time frame to respond to requests by data subjects and a duty to explain immediately when a request by a data subject is denied.[164] When controllers develop new purposes for data use, the data subject must be informed.[165] In some situations, it is difficult or impossible for controllers to contact data subjects. Therefore, the transparency obligations are eliminated in certain situations, for instance when transparency obligations prove impossible or would involve a disproportionate effort.[166]

Fourth, the GDPR commands data users to incorporate protections into the technical design of services with data protection 'by design' and 'by default.' Data protection by design broadly means that privacy rules are implemented in the technical infrastructure, for example pseudonymization and data minimization.[167] Data protection by default indicates that privacy-enhancing choices are made the default in the technical infrastructure, while data subjects can change the default. Data protection by default is intended to prevent services from selecting settings that expose personal data to many other people, for instance, when social media profiles are set to public.[168]

Fifth, the GDPR requires governmental organizations and private organizations that either process personal data on a large scale or process sensitive personal data to appoint a Data Protection Officer (DPO).[169] The DPO occupies a protected, ombudsman-like position to oversee controllers and processors.[170] Many U.S. companies involved in internet marketing will have to have DPOs as they are deemed to process personal data for 'systematic monitoring of data subjects on a large scale'.[171]

The DPO will be a foreign and difficult concept for most U.S. businesses. The closest analogy is to a labor union negotiator. The DPO has to be fully independent,[172] and thus shielded from retaliation when advising the controller or the processor of their obligations. They are tasked with monitoring compliance with the GDPR within the organization; advising the organization on how to become fully GDPR-compliant; cooperating with the Data Protection Authority; and acting as the contact point for both the Data Protection Authority and the data subjects.[173] The DPOs can be seen as the troops on the ground of the Data Protection Authorities, but appointed and paid for by the data controllers themselves.

---

[161]GDPR art 24(2).
[162]GDPR art 12–14.
[163]GDPR art 12.
[164]GDPR art 12(2) and 12(4).
[165]GDPR art 13–14.
[166]GDPR art 14.
[167]GDPR art 25(1).
[168]GDPR art 25(2).
[169]See GDPR art 37–39.
[170]See also GDPR rec 97.
[171]See GDPR art 37(1)(b).
[172]GDPR art 38.
[173]GDPR art 39.

The DPO adds a dynamic that will be difficult for companies to navigate: on one hand, a company might take on the expense of an internal DPO, one that knows the enterprise thoroughly, but also one that is not easy to terminate. On the other, a consultant DPO might not know the enterprise well and have incentives to both interpret the GDPR harshly and document company non-compliance with advice, because an outside DPO might be concerned about saving face with the regulator and other clients.

Sixth, the GDPR expands on the Directive with developing 'Data Protection Impact Assessments' (DPIA) for kinds of processing considered high risk.[174] Whether there is a high risk should be assessed by the data controller itself.[175] The GDPR assumes high risk in three cases. First, when a systematic and extensive evaluation of personal aspects is based on automated processing or profiling and leads to decisions that significantly affect people. An example might be a bank that processes personal data to decide, automatically, whether people can borrow money. A second example of high-risk processing is when special categories of data are processed on a large scale. For instance, when a large hospital processes patients' genetic and health data.[176] Third, when systematically monitoring a publicly accessible area on a large scale.[177] For example, when a company follows people's movements in an airport using Wi-Fi tracking. When a data controller wrongfully assesses a processing operation not to pose a high risk and thus abstains from doing a DPIA, this will qualify as a violation of the GDPR in and for itself.[178]

The DPIA has to describe how and why the data will be processed. But the DPIA must also make some judgment calls on processing – the necessity and proportionality of the processing operations, the risks to the rights and freedoms of data subjects, and finally the measures envisaged to address the risks. If the DPO determines that the processing is high risk despite compensating safeguards, her or she must inform the data protection authority. And where the Data Protection Authority thinks that the intended processing would not comply with the GDPR, it shall give advice on how to mitigate the risks.[179]

Seventh, the GDPR requires data controllers to ensure that organizational safeguards are implemented.[180] Such may include a clean desk policy, restricting the access of personnel to databases and logging access to databases by personnel, in order to validate ex post whether access was indeed necessary and proportionate.

Eighth, the GDPR commands that controllers and processors implement technical safeguards keyed to nature, scope, context, and purposes of the processing as well as the risks of varying likelihood and severity for the rights and freedoms of individuals.[181] Such measures may, for example, include the encryption of data and securing databases with double or triple authentication measures. The GDPR conception of information security incorporates confidentiality, integrity, availability, as well as an interest in system resilience. Yet what measures are required is unclear, as the GDPR both signals a 'state of the art' standard, but tempers the standard with a consideration of the costs involved.[182] It is clear that sensitive data require more security.[183]

---

[174]GDPR art 35(1).
[175]See also Article 29 Working Party, 'Opinion 04/2013 on the Data Protection Impact Assessment Template for Smart Grid and Smart Metering Systems ("DPIA Template") prepared by Expert Group 2 of the Commission's Smart Grid Task Force', 00678/13/EN, WP205, 22 April 2013.
[176]See Article 29 Working Party, 'Guidelines on Data Protection Impact Assessment (DPIA) and determining whether processing is 'likely to result in a high risk' for the purposes of Regulation 2016/679', WP248, 4 April 2017.
[177]GDPR art 35(1)–35(3); GDPR recs 89–94.
[178]GDPR art 83(4)(a).
[179]GDPR art 36.
[180]GDPR art 32.
[181]ibid.

Ninth, there is a personal data breach notification obligation – one much broader than the U.S. approach focusing on just Social Security Numbers, drivers license numbers, and financial account identifiers. The GDPR defines a personal data breach as 'a breach of security leading to the accidental or unlawful destruction, loss, alteration, unauthorised disclosure of, or access to, *personal data* transmitted, stored or otherwise processed'.[184] Thus, under the GDPR, many events currently classified by U.S. companies as security 'incidents' will now be reportable personal data breaches. For instance, if a company loses its customer database of emails, contact details, and the like, it need not be reported under U.S. law but must be under the GDPR.

When a personal data breach occurs, the controller must communicate this to the Data Protection Authority, on an uncomfortably quick time scale—the GDPR requires breach notice to the Data Protection Authority in just 72 hours after its discovery. Notice to the Data Protection Authority is not required if data breach is unlikely to result in a risk to the rights and freedoms of natural persons.[185] If a data breach is likely to result in a high risk for the rights and freedoms of data subjects, the controller must communicate the breach to the data subjects too.[186] Controllers can avoid notification when encryption or other techniques reduce the risk to individuals.[187] Also, where individual notification is disproportionally costly, the controller must inform data subjects through a public communication or similar measure.[188] Controllers must document all data breaches, even the ones that they did not have to notify.[189]

Taken together, these responsibilities require controllers and processors to document compliance, noncompliance, and failures in the form of data breaches. It's apparent that most companies will never be in perfect compliance with the GDPR, and that Data Protection Authorities will have an easier time investigating and policing deviations.

## 6. Rights of the data subject

Europeans enjoy unparalleled data subject rights; they can access, rectify, and erase personal data, and have the right to object to, or restrict, processing.[190] These rights existed in the Directive and have their roots in constitutional instruments, but the GDPR defines them in more detail. In short, the GDPR details seven rights for data subjects: the right to access; to data portability; to rectify data; to stop processing; to object; to erase data; and to resist profiling.

---

[182]GDPR art 32. See on the notion of risk GDPR rec 76.
[183]See on security requirements also: Joined Cases C-293/12 and C-594/12 *Digital Rights Ireland Ltd v Minister for Communications, Marine and Natural Resources and Others and Kärntner Landesregierung and Others* [2014] ECLI:EU:C:2014:238, par 66; A Arnbak and FJ Zuiderveen Borgesius, 'New data security requirements and the proceduralization of mass surveillance law after the European Data Retention Case', Amsterdam Law School Research Paper No. 2015-41.
[184]GDPR art 4(12). See also: Article 29 Working Party, 'Opinion 03/2014 on Personal Data Breach Notification', 693/14/EN, WP 213, 25 March 2014. Article 29 Working Party, 'Opinion 06/2012 on the draft Commission Decision on the measures applicable to the notification of personal data breaches under Directive 2002/58/EC on privacy and electronic communications', 01119/13/EN, WP197, 12 July 2012.
[185]GDPR art 33.
[186]GDPR art 34.
[187]See also GDPR rec 83 and 85–88.
[188]See further Article 29 Working Party, 'Opinion 03/2014 on Personal Data Breach Notification', 693/14/EN, WP 213, 25 March 2014.
[189]GDPR art 32(5).
[190]The rights are not limited to European citizens. See GDPR art 3 on the territorial scope; the nationality of the data subject does not play a role for the scope of the GDPR.

The Charter of Fundamental Rights of the European Union grants all people the right of access to their personal data.[191] The GDPR specifies that data subjects have the right to learn whether controllers are processing their data,[192] and if so, to information about the processing purposes, the categories of personal data, the storage period, the recipients of the data, where the controller obtained the data. The data also subject has the right to obtain a copy of his or her personal data.[193]

Portability is a second right, one included to increase control over data and perhaps competition for data-intensive services.[194] The right to data portability seems to be inspired by number portability, which grants consumers a right to maintain their telephone numbers when changing providers.[195] Under the GDPR's right to data portability, data subjects can take their data from one platform, for example Facebook, to another platform.[196] This right to data portability only applies to personal data that data subjects provided themselves, and the right only applies when either consent or contractual necessity is the legal basis for processing.[197] The data controller must provide such data to the data subject in a structured, commonly used, and machine-readable format.[198]

Third, the GDPR makes specific the right to rectify inaccurate personal data, another right established in the Charter.[199] Americans are familiar with rectification in the form of credit reporting, where individuals can correct inaccurate information and even add supplementary statements to make a record more complete.

Fourth, data subjects can restrict processing, they can object to processing, and exercise the right to erasure. Restriction of processing is a new concept in the GDPR. The GDPR defines 'restriction of processing' as 'the marking of stored personal data with the aim of limiting their processing in the future.'[200] A restriction on processing could be seen as a processing pause, while details about the fairness and lawfulness of a processing activity are examined.[201] In U.S. law the closest analogy would be a consumer's right to have a derogatory and allegedly inaccurate data point removed from a credit report, while the consumer reporting agency investigates its veracity. Data subjects can demand restriction of processing because of inaccuracy, because processing is unlawful, or when the data subject objects to processing based on legitimate interests of the controller (while the balancing process occurs).[202]

---

[191]Charter art 8(2). See also GDPR recs 60–64.
[192]GDPR art 15(1).
[193]GDPR art 15(3). For any further copies requested by the data subject, the controller may charge a reasonable fee based on administrative costs. GDPR art 15(3). The right to obtain a copy must not adversely affect the rights and freedoms of others. GDPR art 15(4). If the data subject makes the request by electronic means, in principle the controller must provide the data in a commonly used electronic form. GDPR art 15(3) GDPR. Under certain circumstances, data controllers must inform data subjects about access, rectification, and erasure rights (GDPR art 13(2)(b) and GDPR art 14(2)(c) GDPR).
[194]GDPR art 20; rec 68.
[195]The Universal Services Directive requires phone companies to offer number portability. Directive 2002/22/EC on universal service and users' rights relating to electronic communications networks and services [2002] OJ L108/51 art 30(1).
[196]GDPR art 20.
[197]GDPR art 20 (1).
[198]GDPR art 20(1). The preamble adds that the data portability right does not oblige 'controllers to adopt or maintain processing systems which are technically compatible.' GDPR rec 68.
[199]GDPR art 16 GDPR; Charter art 8(2).
[200]GDPR art 4(3). See also GDPR rec 65.
[201]Where processing has been restricted, the personal data may only be processed with the data subject's consent, for legal reasons, or for an important public interest. GDPR art 18(2).

Fifth, data subjects may also stop processing by objecting in certain circumstances.[203] When controllers justify processing with the legitimate interest or public interest rationales,[204] data subjects may object, stopping processing and triggering a balancing inquiry. After objection, the controller can continue processing if it can demonstrate compelling legitimate grounds for the processing that override the interests of the data subject, or for exercising legal claims.[205] For example, a state body that processes personal data can override a data subject objection for reasons of public interest.[206] As noted previously, the GDPR has a special regime for direct marketing: people have an absolute right to object to (opt out of) direct marketing.[207]

Sixth, the GDPR articulates the 'right to erasure ("right to be forgotten")'.[208] An erasure right also appeared in the Directive. The GDPR adds the phrase 'right to be forgotten' – an unfortunate phrase, because people cannot require others to forget them. The right to erasure is widely derided despite having many antecedents in U.S. law.[209] Erasure is a broader right than in U.S. Law.

Roughly summarized, a data subject has a right to erasure when he or she successfully exercises the right to object, when the personal data were unlawfully processed, should be erased because of a legal obligation, or are no longer necessary in relation to the processing purposes.[210] The right to erasure also applies when the data subject withdraws consent.[211] Hence, the right to erasure provides another rationale why consent is a dangerous basis for processing data.

The preamble says that the erasure right is especially relevant 'where the data subject has given his or her consent as a child and is not fully aware of the risks involved by the processing, and later wants to remove such personal data, especially on the internet.'[212] An example could be an old profile on a social networking site that somebody made when young, including embarrassing pictures. One might want to have the profile deleted. In many cases, personal data will be spread from the original source all over the internet. Therefore, the GDPR requires the controller to make a reasonable effort to erase the personal data elsewhere on the internet, taking account of available technology and the costs of implementation.[213]

The right to erasure does not apply when erasure conflicts with the freedom of expression and the latter interest prevails.[214] Striking this balance is notoriously difficult.

---

The Court of Justice of the European Union decided in its Google Spain judgment that data subjects in Europe have, under certain conditions, the right to have search results for their name delisted from the search results, even where the search result was sourced from a newspaper.[215]

Finally, the GDPR regulates profiling and computerized decision-making processes.[216] Profiling is processing of personal data

> to evaluate certain personal aspects relating to a natural person, in particular to analyse or predict aspects concerning that natural person's performance at work, economic situation, health, personal preferences, interests, reliability, behaviour, location or movements.[217]

This definition covers behavioral advertising and credit scoring.[218]

Under this definition, businesses profile all the time. The GDPR intervenes by subjecting certain types of, *fully automated* profiling to substantive and procedural rules. If profiling is fully automated and can produce legal effects or similarly significantly affect him or her,[219] it is prohibited unless an exception applies.

Automated decisions can be taken if it is (a) necessary for a contract between the data subject and the data controller; (b) if it is authorized by a law which includes suitable safeguards for the data subject's interests; or (c) is based on the data subject's explicit consent.[220] In situation (a) (contract) and (c) (consent), the data controller must implement suitable measures to safeguard the data subject's interests, at least the right to obtain human intervention on the part of the controller, to express his or her point of view, and to contest the decision.[221]

The path to compliance is straightforward. Supposed Jane applies for insurance from John. John can use software to decide whether Jane is insurable, and under what terms, because this is necessary for a contract. Jane's interests can be satisfied if John operates a phone system, where Jane can call and ask a human to reconsider the automated decision that shaped the insurance contract.

In real life, much of automated decision-making supplements human judgment, and these systems appear to escape the prohibition. For instance, in credit scoring, the evaluation (in theory) assists the sales manager in deciding whether to grant credit. The GDPR's provision only applies to a 'decision based *solely* on automated processing.'[222] The predecessor of the GDPR's automated decision provision from the Directive has not been applied much in practice; it has remained a 'dead letter.'[223] It seems plausible that the GDPR's provision will not be applied much in practice either.[224]

---

[215] Case C-131/12 *Google Spain SL, Google Inc. v Agencia Española de Protección de Datos (AEPD), Mario Costeja González* [2014] ECLI:EU:C:2014:317. See also: S Kulk and FJ Zuiderveen Borgesius, 'Privacy, Freedom of Expression, and the Right to Be Forgotten in Europe,' in J Polonetsky, O Tene, and E Selinger (eds) *Cambridge Handbook of Consumer Privacy* (Cambridge University Press, 2018).
[216] GDPR art 22. The provision is based on Article 15 of the Directive, which was inspired by French law. See on that old provision: LA Bygrave, 'Minding the Machine: Article 15 of the EC Data Protection Directive and Automated Profiling' (2001) 17 Computer Law & Security Report 17.
[217] GDPR art 4(4).
[218] See GDPR recs 24 and 71.
[219] GDPR art 22(1). The provision says that a person has 'the right not to be subject to' certain decisions. But many scholars assume that this right implies a prohibition (with exceptions) of such decisions. See, e.g. D Korff, 'Comments on Selected Topics in the Draft EU Data Protection Regulation' (17 September 2012) <http://ssrn.com/abstract=2150145>; P De Hert and S Gutwirth, 'Regulating profiling in a democratic constitutional state,' in M Hildebrandt and S Gutwirth (eds), *Profiling the European Citizen* (Springer, 2008); S Wachter, B Mittelstadt and L Floridi, 'Why a Right to Explanation of Automated Decision-Making does Not Exist in the General Data Protection Regulation' (2017) 7(2) International Data Privacy Law 76; FJ Zuiderveen Borgesius, Improving privacy Protection in the Area of Behavioural Targeting (Kluwer Law International, 2015) 283–93. See also L Edwards and M Veale, 'Slave to the Algorithm: Why a Right to an Explanation is Probably Not the Remedy You are Looking for' (2017) 16 Duke Law & Technology Review 18.
[220] GDPR art 22(2).
[221] Exception (c) is new in the GDPR.

It is prohibited to base fully automated decisions with legal or similar effects on special categories of data, unless the controller gains explicit consent or processing is necessary for reasons of substantial public interest.[225] However, it is unclear to what extent this prohibition applies when an automated decision is based on non-sensitive personal data (such as postal code) that function as a proxy for special categories of data (such as race).

The GDPR also subjects automated decision making to new transparency and accuracy requirements. Under certain circumstances, the data subject has the right to obtain 'meaningful information about the logic involved, as well as the significance and the envisaged consequences of such processing for the data subject.'[226] New in the GDPR is an explicit requirement for controllers using profiling to minimize the risk of errors:

> the controller should use appropriate mathematical or statistical procedures for the profiling, implement technical and organisational measures appropriate to ensure, in particular, that factors which result in inaccuracies in personal data are corrected and the risk of errors is minimised.[227]

In conclusion, regarding data subject's rights, the GDPR follows broadly the same logic as the Directive. But the GDPR introduces new rights (to restrict processing and to data portability) and gives more details regarding the data subject's rights.

The aim of the GDPR is primarily to close the gap between the legal principles – most of which were established in the Directive – and practice. The GDPR achieves this by both specifying additional obligations for data controllers and by making more explicit data subjects' rights. These obligations try to ensure that data controllers implement and execute the various data protection principles. But the main new feature of the GDPR is that the GDPR takes enforcement much more seriously than the Directive. Enforcement is discussed in the next section.

## 7. Enforcement

The GDPR invests heavily in compliance and enforcement. We discuss three key enforcement issues here: first, who can sue, and the amount of sanction imposed, is dramatically expanded under the GDPR. The GDPR enables class-action-like activities in Europe, and fines that approach those established in competition law. Second, the tasks and powers of Data Protection Authorities have expanded considerably. For businesses, this is a silver lining – the tasks burden authorities with functions and necessarily will reduce enforcement efforts. On the other hand, the requirement that Data Protection Authorities be independent means a relative immunity from political pressure, similar to what judges and independent agencies enjoy in the U.S. There are also many Data Protection Authorities; they can share investigatory information and cooperate in investigations, creating scary dynamics for companies that are similar to multi-state efforts by the attorneys general in the U.S. Third, the European Commission and an advisory board have been granted powers for standard setting. We discuss each point below.

---

[222]GDPR art 22(1). The Article 29 Working Party interprets this phrase more generously for the data subject. Article 29 Working Party, 'Guidelines on Automated individual decision-making and profiling for the purposes of Regulation 2016/679' (WP251rev.01), 6 February 2018.
[223]D Korff, 'Comments on Selected Topics in the Draft EU Data Protection Regulation' (17 September 2012) <http://ssrn.com/abstract=2150145>.
[224]See I Mendoza and LA Bygrave, 'The Right Not to be Subject to Automated Decisions Based on Profiling' (2017) <https://ssrn.com/abstract=2964855>.
[225]GDPR art 22(4).
[226]GDPR art 12(2)(f).
[227]GDPR rec 71.

The Directive was plagued by ineffective sanctions. The Directive left fines and other remedies to individual member states. Some countries implemented the Directive with maximum fines of a couple of thousand euros – so low to be completely inconsequential to many businesses. For instance, the CNIL levied its maximum fine against Facebook in 2017 for tracking users for advertising purposes. The fine was 150,000.[228]

The GDPR brings about a change on three points: sanctions, remedies and liability. The changes with respect to sanctions are the most spectacular. With regard to the right to compensation and liability, the GDPR has little new. As for sanctions, there are two levels of fines, based on seriousness of the violation. Less serious violations can trigger administrative fines up to ten million euro or, in the case of an undertaking, up to 2% of the total worldwide annual 'turnover' (net – not profit) of the preceding financial year, whichever is higher. The 2% sanction is triggered when the data controller violates the rules on, inter alia, consent given by children, data protection by design and by default, the keeping of documents and records on the data processing activities, notifying a data breach, and data protection impact assessments.[229]

More serious violations can trigger administrative fines of up to 20 million euro or, in the case of an undertaking, up to 4% of the total worldwide annual turnover of the preceding financial year, whichever is higher. The 4% sanction is triggered, inter alia, in case of a violation of the data minimization principle, the purpose limitation principle, the accuracy principle, the integrity and confidentiality principle, the rights of data subjects (such as the right to be informed and the right to correct or erase data), the transparency obligations, and the rules on trans-border data flows.[230]

To place this in perspective, Facebook's 2017 fine of €150,000 could rise to between €800 mm to €1,6bn under the GDPR. At least in theory. Similarly, in theory, American businesses are subject to crippling, trillion-dollar liability for aggregated statutory fines associated with laws such as the Fair Credit Reporting Act. For instance, if Equifax were to be fined for failing to secure 140 million credit records under the FCRA's minimum damage amount, it would face a 12-figure fine. However, no U.S. court imposes such awards. In fact, the largest FCRA fine appears to have been just $18 million, and it was reduced on appeal.[231] Furthermore, there is no example of a U.S. business failing because of a privacy fine.

The European reality will be similar. The GDPR dictates that fines be proportional and keyed to the seriousness of the offense.[232] We expect that the first fines under the GDPR will fall far short of 9 and 10-figure amounts until Data Protection Authorities build a set of cases with egregious behavior. This might take decades. In the meantime, the expanded notion of security breaches will cause more pain than fines. Already, there has been a dramatic uptick in breach reports to regulators. Because a much broader scope of security incident needs to be reported to regulators, we think breach notification will be the primary form of 'punishment' that regulated entities experience.

---

[228]CNIL, 'Facebook sanctioned for several breaches of the French Data Protection Act,' May 16, 2017 <https://www.cnil.fr/en/facebook-sanctioned-several-breaches-french-data-protection-act>.
[229]GDPR art 83(4).
[230]GDPR art 83(5) and 83(6).
[231]Andrew Scurria, 'Equifax Gets Record-Setting $18M FCRA Verdict Slashed,' *Law360* (January 30, 2014).
[232]GDPR art 83(1) and 83(2).

Turning to remedies, here too the GDPR provides new tools that presage more vigorous enforcement than under the Directive. On the most basic level, individuals can file complaints with Data Protection Authorities. If dissatisfied with the authority's decision, the individual can seek judicial remedy.[233]

In the U.S., individuals have judicial remedies on paper, but the standing doctrine has undermined the enforceability of many statutory privacy laws. To address this problem, the GDPR specifies that each data subject has the right to an effective judicial remedy if they consider that their rights have been infringed as a result of the processing of their personal data.[234] In fact, to overcome the collective action and economic barriers to individual suit, the GDPR creates a class-action-like mechanism. Data subjects can have their complaint brought by a non-profit devoted to protecting rights.[235] Austrian lawyer Max Schrems, the litigant responsible for invalidation of the U.S.-EU Safe Harbor and many headaches for the U.S. business community, has already created such an entity using crowdfunding. On the day the GDPR became enforceable, Schrems' group lodged complaints against Google, Instagram, WhatsApp, and Facebook.[236]

A second enforcement shift comes in the powers of Data Protection Authorities. Such authorities have broad powers to investigate, to intervene and even halt data processing, and to bring legal proceedings. Table 1 summarizes the powers of Data Protection Authorities.

While Data Protection Authorities have more enforcement powers under the GDPR, they also have many duties that will consume resources and prevent all but the most focused authorities from becoming aggressive enforcers. Table 2 summarizes the tasks of Data Protection Authorities.

Data Protection Authorities have been understaffed, under-resourced, and sometimes have not enjoyed adequate independence. The CJEU has decided in a few cases that certain member states did not sufficiently safeguard the independence of its Data Protection Authority.[237] The GDPR strengthens supervisory independence.[238]

In the U.S. the Federal Trade Commission takes consumer complaints but does not adjudicate them. Instead, the FTC sets policy by selecting consequential cases, sometimes completely untethered from the complaints received. Data Protection Authorities, on the other hand, are obliged to hear complaints, which will cost them a lot of time.[239]

As explained in Section 2, the Data Protection Directive from 1995 ordered Member States to pass implementing privacy legislation. Some states passed weak laws and signaled a business-friendly environment with weak-wristed Data Protection Authorities.

---

[233]GDPR art 77 and 78.
[234]GDPR art 79.
[235]GDPR art 80.
[236]See <https://noyb.eu>
[237]See for instance: Case C-288/12 Commission v Hungary ECLI:EU:C:2014:237 8 April 2014; Case C-614/10, CJEU (Grand Chamber) European Commission v Republic of Austria, 16 October 2012.
[238]GDPR arts 51–54. See further GDPR recs 117–124.
[239]See GDPR art 57(1)(f).

**Table 1.** Powers of data protection authorities, summary[a].

| Investigatory powers | Corrective powers | Authorization and advisory powers |
| --- | --- | --- |
| a. Order the controller and the processor to provide information to Data Protection Authorities (DPAs) | a. Issue warnings to a controller or processor that intended processing operations are likely to infringe the GDPR | a. Advise the controller (prior consultation procedure) |
| b. Carry out investigations in the form of data protection audits | b. Issue reprimands to a controller or a processor where processing operations have infringed the GDPR | b. Issue opinions on data protection issues, to the national government or to other institutions and to the public |
| c. Carry out a review on certifications issued | c. Order the controller or processor to comply with the data subject's requests to exercise his or her rights | c. Authorize processing, if Member State law requires such prior authorization |
| d. Notify the controller or the processor of an alleged infringement of the GDPR | d. Order the controller or processor to bring processing operations into compliance with the GDPR | d. Assess and approve draft codes of conduct |
| e. Obtain, from the controller and the processor, access to all personal data and necessary information | e. Order the controller to communicate a personal data breach to the data subject | e. Accredit certification bodies |
| f. Obtain access to premises of the controller and the processor, including to any data processing equipment and means | f. Impose a temporary or definitive ban on processing | f. Issue certifications and approve criteria of certification |
| | g. Order the rectification, restriction or erasure of data, and the notification of such actions to recipients of the data | g. Adopt standard data protection clauses |
| | h. Withdraw a certification or order the certification body to withdraw a certification issued, or to order the certification body not to issue certification | h. Authorize contractual clauses |
| | i. Impose an administrative fine | i. Authorize administrative agreements |
| | j. Order the suspension of data flows to a recipient in a third country | j. Approve binding corporate rules |

[a]GDPR art 58.

Many companies chose to locate their European headquarters in countries with a low regulatory burden or with underequipped Data Protection Authorities.

The GDPR specifies which Data Protection Authority is the main regulator for a company. A company's 'lead supervisory authority'[240] is based on the 'main establishment' of the controller or processor.[241] The 'main establishment' is the place of a controller's or processor's central administration, or the place where the decisions on the purposes and means of the processing (with respect to processors, where the main processing takes place) of personal data are taken.[242] Although there is a lead authority, the GDPR specifies that authorities can cooperate, have joint operations,[243] and share information,[244] making it risky for companies to communicate different messages.

---

[240]GDPR recs 125–128.
[241]GDPR art 56.

**Table 2.** Tasks of data protection authorities, summary[a].

| | |
|---|---|
| a. Monitor and enforce the application of the GDPR | b. Promote public awareness of the risks, rules, safeguards and rights in relation to personal data processing, especially when children are affected |
| c. Advise the national government and other institutions on legislative measures relating to processing personal data | d. Promote the awareness of controllers and processors of their obligations |
| e. Upon request, provide information to data subjects concerning the exercise of their rights | f. Deal with complaints by data subjects |
| g. Cooperate with other Data Protection Authorities | h. Conduct investigations on the application of the GDPR |
| i. Monitor relevant developments, in particular the development of information and communication technologies and commercial practices | j. Adopt standard contractual clauses |
| k. Maintain a list of processing operations that require a data protection impact assessment | l. Advise in relation to prior consultation regarding high risk processing |
| m. Encourage the drawing up of codes of conduct and assess and possibly approve such codes | n. Encourage the establishment of data protection certification mechanisms and data protection seals and marks, and possibly approve the criteria of certification |
| o. Carry out a periodic review of certifications | p. Publish the criteria for accreditation of a body for monitoring codes of conduct and of a certification body |
| q. Conduct the accreditation of a body for monitoring codes of conduct and of a certification body | r. Authorize contractual clauses and provisions |
| s. Approve binding corporate rules | t. Contribute to the activities of the European Data Protection Board |
| u. Keep internal records of breaches of the GDPR and of measures taken, in particular warnings issued and sanctions imposed | v. Fulfill any other tasks related to the protection of personal data |

[a]GDPR art 57.

Those familiar with the Directive know that the Article 29 Working Party was a body that made non-binding but persuasive recommendations on interpretations of the law and that it postured in favor of more data protection rights.[245] The GDPR strengthens the voice of Data Protection Authorities on a Europe-wide level through the transformation of the Working Party into the European Data Protection Board (EDPB). The EDPB is composed of the head of one Data Protection Authority of each Member State and of the European Data Protection Supervisor.[246]

---

[242]GDPR art 4(16). See further GDPR rec 36.
[243]GDPR art 62.
[244]GDPR art 66.
[245]Directive art 29.
[246]GDPR art 68. See on its functioning also: GDPR recs 139–143.

**Table 3.** Tasks of the European data protection board, summary[a].

| | |
|---|---|
| a. Monitor and ensure the correct application of the GDPR | b. Advise the European Commission on any personal data issue |
| c. Advise the European Commission on the exchange of information between controllers and Data Protection Authorities for binding corporate rules | d. Issue guidelines, recommendations, and best practices on procedures for deleting links, copies or replications of personal data (right to be forgotten) |
| e. Examine any question covering the application of the GDPR and issue guidelines etc. to encourage consistent application | f. Issue guidelines etc. for specifying the criteria and conditions for decisions based on profiling |
| g. Issue guidelines etc. regarding data breaches and their notification | h. Issue guidelines etc. regarding the circumstances in which a data breach is likely to result in a high risk |
| i. Issue guidelines etc. regarding binding corporate rules | j. Issue guidelines etc. regarding data transfers to non-EU countries |
| k. Draw up guidelines for DPAs regarding the powers of DPAs and regarding fines | l. Review the practical application of the guidelines etc. regarding the consistent application of the GDPR and regarding profiling |
| m. Issue guidelines etc. for establishing common procedures for reporting by individuals of GDPR infringements | n. Encourage the drawing-up of codes of conduct and the establishment of data protection certification mechanisms and data protection seals and marks |
| o. Carry out the accreditation of certification bodies and maintain a public register of accredited bodies | p. Specify the requirements regarding the accreditation of certification bodies |
| q. Advise the European Commission on certification requirements | r. Advise the European Commission on icons (for transparency) |
| s. Advise the Commission on the assessment of the adequacy of non-EU countries | t. Issue opinions and adopt decisions to ensure the consistent application of the GDPR in the EU |
| u. Promote the cooperation between DPAs | v. Promote common training programs and facilitate personnel exchanges between DPAs |
| w. Promote the exchange of knowledge with DPAs and similar bodies worldwide | x. Issue opinions on codes of conduct drawn up at EU level |
| y. Maintain a public register of decisions taken by DPAs and courts that are relevant for the consistent application of the GDPR | |

[a]GDPR art 70.

The tasks of the EDPB include advising national Data Protection Authorities,[247] advising the European Commission;[248] and issuing guidelines, recommendations, and best practices.[249] The EDPB's tasks and powers are summarized in Table 3.

## 8. Conclusion

This paper introduced the normative foundations, attributes, and strategic approach to regulating personal data advanced by the European Union's GDPR. We explained the genesis of the GDPR; described the GDPR's approach and provisions; and made predictions about the GDPR's short and medium-term implications. We showed that the GDPR is best understood as an extension and refinement of existing requirements imposed by the 1995 Data Protection Directive. The core of data protection law proves to be remarkably stable. Data protection's core principles, comparable to the Fair Information Principles, are retained in the GDPR.

---

[247]GDPR art 70(1)(a).
[248]GDPR art 70(1)(b), 70(1)(c), and 70(1)(q) – 70(1)(s).
[249]GDPR art 70(1)(d) – 70(1)(m).

However, the GDPR also brings significant changes. The GDPR brings personal data into a complex, detailed, and protective regulatory regime, which will have profound implications. For example, the GDPR encourages companies to think carefully about their personal data practices. The GDPR attempts to make companies take privacy seriously. The GDPR encourages companies to vet service providers for their compliance with the rules Another innovation is that the GDPR elevates privacy officials within organizations. Furthermore, the GDPR is skeptical of the legal tool of 'informed consent': in some circumstances, organizations cannot rely on consent. And the GDPR emphasizes the importance of accurate data and grants people the right to access and correct data. We foresee that, in the private sector, first party relationships with individuals become more important, to the detriment of third party relationships. We also expect an extended tussle between Data Protection Authorities and large technology companies about the interpretation of the GDPR.

The main disadvantage of the GDPR is its length and complexity: 99 detailed provisions. Whether the GDPR will actually improve fairness and respect for fundamental rights can, of course, only be assessed when it has been applicable for some time.

Meanwhile, the European Commission has started its next privacy project. The Commission has published a proposal for an ePrivacy Regulation, which should replace the ePrivacy Directive. The proposal includes rules to protect confidentiality of communications on the internet, and rules regarding cookies and online tracking.[250]

The rules for fair processing of personal data will never be finished. Just like in consumer protection law or environmental law, the rules will have to be updated and amended continually, to adapt to new circumstances. In conclusion, the GDPR signals a new phase in privacy law, and will influence policy worldwide.

\* \* \*

## Disclosure statement

## Funding

This work was supported by the European Union, with an EU Marie Curie individual grant [grant number 748514, PROFILE].

---

[250]See FJ Zuiderveen Borgesius, J Van Hoboken, K Irion, and M Rozendaal, 'An Assessment of the Commission's Proposal on Privacy and Electronic Communications,' Directorate-General for Internal Policies, Policy Department C: Citizen's Rights and Constitutional Affairs, June 2017, <https://ssrn.com/abstract=2982290>.